\documentclass[12pt,preprint]{aastex}

\shorttitle{MUSYC Wide NIR Survey}
\shortauthors{Blanc et al.}

\begin{document}

\title{THE MULTIWAVELENGTH SURVEY BY YALE-CHILE (MUSYC): Wide K-band Imaging, Photometric Catalogs, Clustering and Physical Properties of Galaxies at z$\sim$2}

\author{Guillermo A. Blanc,\altaffilmark{1,2} Paulina Lira,\altaffilmark{1} L. Felipe Barrientos,\altaffilmark{3} Paula Aguirre,\altaffilmark{3} Harold Francke,\altaffilmark{1} Edward N. Taylor,\altaffilmark{6} Ryan Quadri,\altaffilmark{4} Danilo Marchesini,\altaffilmark{4} Leopoldo Infante,\altaffilmark{3} Eric Gawiser,\altaffilmark{5} Patrick B. Hall,\altaffilmark{7} Jon P. Willis,\altaffilmark{8} David Herrera,\altaffilmark{9} Jos\'e Maza\altaffilmark{1} for the MUSYC Colaboration}

\email{gblancm@astro.as.utexas.edu}

\altaffiltext{1}{Departamento de Astronom\'ia, Universidad de Chile, Chile}
\altaffiltext{2}{Astronomy Department, University of Texas at Austin, USA}
\altaffiltext{3}{Departamento de Astronom\'ia, Pontificia Universidad Cat\'olica de Chile, Chile}
\altaffiltext{4}{Astronomy Department, Yale University, USA}
\altaffiltext{5}{Department of Physics and Astronomy, Rutgers University, USA}
\altaffiltext{6}{Leiden Observatory, Leiden University, Netherlands}
\altaffiltext{7}{Department of Physics and Astronomy, York University, Canada}
\altaffiltext{8}{Department of Physics and Astronomy, University of Victoria, Canada}
\altaffiltext{9}{National Optical Astronomy Observatory, USA}

\begin{abstract}

We present K-band imaging of two $\sim 30' \times 30'$ fields covered by the MUSYC Wide NIR Survey. The 1030 and 1255 fields were imaged with ISPI on the 4m Blanco telescope at CTIO to a $5 \sigma$ point-source limiting depth of $K \sim 20$ (Vega). Combining this data with the MUSYC Optical UBVRIz imaging, we created multi-band K-selected source catalogs for both fields. These catalogs, together with the MUSYC K-band catalog of the ECDF-S field, were used to select $K<20$ BzK galaxies over an area of 0.71 deg$^2$. This is the largest area ever surveyed for BzK galaxies. We present number counts, redshift distributions and stellar masses for our sample of 3261 BzK galaxies (2502 star-forming (sBzK) and 759 passively evolving (pBzK)), as well as reddening and star formation rate estimates for the star-forming BzK systems. We also present 2-point angular correlation functions and spatial correlation lengths for both sBzK and pBzK galaxies and show that previous estimates of the correlation function of these galaxies were affected by cosmic variance due to the small areas surveyed. We have measured correlation lengths $r_0$ of $8.89\pm2.03$ Mpc and $10.82\pm1.72$ Mpc for sBzK and pBzK galaxies respectively. This is the first reported measurement of the spatial correlation function of passive BzK galaxies. In the $\Lambda$CDM scenario of galaxy formation, these correlation lengths at $z\sim2$ translate into minimum masses of $\sim 4\times 10^{12}$M$_{\odot}$ and $\sim 9\times 10^{12}$M$_{\odot}$ for the dark matter (DM) halos hosting sBzK and pBzK galaxies respectively. The clustering properties of the galaxies in our sample are consistent with them being the descendants of bright LBG at $z\sim3$, and the progenitors of present-day $>1$L$^{*}$ galaxies.

\end{abstract}


\section{Introduction}

The processes involved in the formation of galaxies and the phenomena that drive their consequent evolution are among the most important and fundamental problems in astronomy. In order to understand these processes we need to study the properties of statistically significant samples of galaxies throughout cosmic time. During the last decade there have been enormous advances in this direction. The advent of 8m class telescopes, a new generation of wide field optical imagers and infrared detectors, and a series of different photometric selection techniques, have provided us with the ability to select large samples of intermediate and high redshift galaxies. Probably the most successful method applied to date in order to build these samples has been the Lyman Break Galaxy (LBG) dropout technique at $z\sim3$ \citep{steidel96} which has recently been extended to lower redshifts ($z\sim2$) using $U_nG\cal{R}$ two-color photometry, the BM/BX galaxy selection \citep{erb03, adelberger04}. Other types of $z>1$ galaxies include narrow-band selected Lyman Alpha Emitters (LAE) (e.g. \cite{hu98, gawiser06b, gronwall07}), Distant Red Galaxies (DRG) selected by having red near-infrared colors $J-K>2.3$ \citep{franx03, vandokkum03}, Extremely Red Objects (ERO) selected by their red optical to near-infrared colors, usually $R-K>5$ \citep{thompson99, cimatti02a} as well as far-IR and sub-mm detected, highly obscured galaxies at high redshift \citep{smail00}.

Another selection technique, of great importance to this work, was introduced by \cite{daddi04} who defined a two color criteria based on BzK photometry of K-selected galaxies, which is capable of identifying both passive (pBzK) and star-forming (sBzK) galaxies in the $1.4<z<2.5$ range, and to distinguish between the two populations. Among the available NIR bands, the K band is the best choice to detect intermediate and high redshift galaxies from the ground. It is the reddest band for which sensitivities and angular resolutions comparable to the ones obtained by ground-based optical imaging can be achieved. Since galaxies can be selected efficiently by their rest-frame optical emission redshifted to the NIR bands, K-selection gives us access to the  ``spectroscopic redshift desert'' ($1.4<z<2$) where galaxies show no distinctive spectral features in their observed optical spectra. One of the main advantages of the BzK technique is that it is almost insensitive to the amount of reddening in the star-forming galaxies. NIR observations allow us to detect and identify distant galaxies in a more unbiased way than optical imaging, since the effects of dust extinction are lower at these wavelengths and the observed light is dominated by evolved stellar populations, making detection more independent of spectral type. NIR luminosity is directly related to the stellar mass of the galaxy. The BzK selection technique provides us with a powerful tool to explore the $z\sim2$ universe in a relatively unbiased way.

Despite all these advances, and the fact that we have converged into a single cosmological model, the assembling and evolution of galaxies remains poorly understood. We still lack a comprehensive picture of galaxy populations at intermediate and high redshifts. All the selection techniques mentioned above only give partial views of the universe at specific epochs, and the biases involved in their selection, as well as the overlapping between different techniques is just starting to be understood \citep{reddy05, vandokkum06, quadri07b, lane07, grazian07}.
In particular, optically selected samples require galaxies to be bright in their rest-frame UV, and are therefore biased against obscured systems in which the UV luminosity of a star-forming galaxy is quenched by the dust content of its ISM, as well as against passively evolving systems with little intrinsic UV emission. Hence, studies based on these samples constrain the behavior and evolution of unobscured star formation activity, more than the actual stellar mass assembly of these systems. At $z\sim2$ the most popular selection techniques used so far include UV selected BM/BX galaxies, DRGs, and BzK galaxies. \cite{reddy05} showed that BM/BX selection misses 40\% of star-forming BzK galaxies when both samples are subject to a common NIR magnitude limit of $K<20$. Recent work by \cite{quadri07b} estimated a larger missing fraction of 60\% to $K<20$, which decreases to a 35\% when going to deeper NIR fluxes ($K<21$). \cite{grazian07} computed an even larger fraction of missed sBzK galaxies of a 62\% at a deeper NIR magnitude limit of $K<22$. In all the mentioned works, BM/BX galaxies completely miss passive BzK galaxies at $z\sim2$. UV based selection techniques perform even more poorly when trying to recover the DRG population. \cite{quadri07b} showed that only 32\% of DRG in the $K<21$ MUSYC Deep NIR sample have colors consistent with the BM/BX/LBG criteria, and only 14\% also fulfill the $\cal{R}$$_{AB}<25.5$ cut used in most ground-based optically selected samples. This is in agreement with the 12\% recovery fraction estimated by \cite{reddy05} to the same NIR magnitude limit and higher than the 3\% fraction computed by \cite{grazian07} for a deeper $K<22$ DRG sample.

On the other hand, the sBzK criteria successfully recovers 80\% of the $z>1.4$, $K<20$ BM/BX galaxies in Reddy's sample, in agreement with the results of \cite{quadri07b} for a $K<21$ sample. This fraction improves to up to a 99\% when going to a fainter NIR limit of $K<22$ \citep{grazian07}. The BzK technique is also much better than rest-frame UV selection techniques at recovering the DRG population. For a bright $K<19.3$ sample of DRG, \cite{lane07} obtained a recovery fraction of 55\% being selected as either sBzK or pBzK. A higher fraction of 65\% was obtained by \cite{reddy05} for $K<21$ DRG, again in agreement whith the results of \cite{quadri07b}. When going to fainter NIR fluxes($K<22$) \cite{grazian07} estimated an almost complete recovery fraction of 99\%. These numbers show that the BzK technique is the most inclusive galaxy selection method at $z\sim2$, mostly because is able to include galaxies that span a wide range of both optical and optical-to-NIR colors.

Previous samples of BzK galaxies include the seminal work by \cite{daddi04} who selected $K<20$ BzK galaxies over an area of 52 arcmin$^2$, a survey by \cite{kong06} who covered 320 arcmin$^2$ to $K<20$ and 920 arcmin$^2$ to a shallower depth of $K=19$, and more recent work by \cite{lane07} who surveyed 0.56 deg$^2$ to $K=20.6$ in the UKIDSS field, and \cite{hayashi07} who selected very faint BzK galaxies ($K=21.3$) over a 180 arcmin$^2$ area in the Subaru Deep Field.

In this work, we present the MUSYC Wide NIR Survey, and the clustering and physical properties of BzK galaxies selected over an area of 0.71 deg$^2$ to a limiting depth of $K=20$. This is the largest area ever surveyed for BzK galaxies, and the largest sample of these objects available to this depth. In Section 2 we introduce the MUSYC survey and present the Wide NIR Survey design. Sections 3 and 4 discuss the observations and data reduction. Source detection, photometry and the procedures used to build the K-selected source catalogs are discussed in Section 5. In Section 6 we explain our Star-Galaxy classification method, and we present K-band selected galaxy number counts in Section 7. We present the details of the BzK selection technique used in this paper in Section 8. The results coming from the analysis of the selected BzK sample, redshift distributions, physical properties and clustering are presented in Section 9. Finally, we summarize our results and present conclusions in Section 10.

All optical magnitudes are given in the AB system, and NIR magnitudes in the Vega system. Throughout the whole paper we adopted the standard cosmological parameters, $H_o=71$ km s$^{-1}$Mpc$^{-1}$ and  $\Omega_M=0.27$, $\Omega_{\Lambda}=0.73$ and $\sigma_8=0.8$. All quantities reported are comoving, so correlation lengths scale as $h_{70}^{-1}$, number densities as $h_{70}^3$ and halo masses as $h_{70}^{-1}$, unless explicitly stated.

\section{The MUSYC Wide NIR Survey}

The MUltiwavelength Survey by Yale Chile (MUSYC) consists of deep optical (UBVRIz) and near-infrared imaging of four $30' \times 30'$ fields. The survey design and optical images are described by \cite{gawiser06a}. The four fields (Extended Hubble Deep Field South (EHDF-S), Extended Chandra Deep Field South (ECDF-S), SDSS1030+05 and Cast1255) are at high galactic latitudes ($\mid\! b\! \mid > 30$) and show low 100 $\mu$m emission and N$_{\rm{H}}$ in order to minimize contamination by stars and the Galaxy ISM.

The near-infrared imaging of the MUSYC fields forms part of two complementary campaigns. The MUSYC Deep NIR Survey imaged four $10' \times 10'$ sub-fields in three NIR bands to typical $5 \sigma$ limiting depths of $J \sim 22.5$, $H \sim 21.5$ and $K \sim 21$. The survey design and results are presented by \cite{quadri07b}. The MUSYC Wide NIR Survey consists of K-band imaging of three full $30' \times 30'$ MUSYC fields (ECDS-S, SDSS1030+05 and Cast1255) to a limiting depth of $K\sim 20$ as well as J and H-band imaging of the ECDF-S.

In this work we present the NIR imaging and K-band selected source catalogs for two fields of the Wide NIR Survey: SDSS1030+05 and Cast1255 (hereafter 1030 and 1255). Table \ref{tbl-1} gives relevant information about the surveyed fields. The wide NIR imaging of the ECDF-S will be presented by \cite{taylor07}. H-band imaging to similar depth of EHDF-S is already available from the Las Campanas Infrared Survey \citep{chen02}.

MUSYC also includes narrow-band 5000\AA\ imaging of the whole square degree to search for Lyman Alpha Emitters and Lyman Alpha Blobs at $z=3.1$ \citep{gawiser06b, blanc08}, mid-infrared from the Spitzer IRAC MUSYC Public Legacy Survey (SIMPLE, \cite{labbe07}), Chandra X-ray imaging \citep{virani06b} of the ECDF-S, XMM-Newton X-ray observations of 1030, and spectroscopic follow up of the four fields \citep{lira08}.

More information about MUSYC, together with publicly available data can be found at {\it http://www.astro.yale.edu/musyc/}.

\section{Observations}

Our K-band images of the 1030 and 1255 fields were taken on the nights of 2003 January 31, 2003 May 7-12, 2004 February 11-14, 2004 April 10-13, 2005 April 14-22, and 2006 April 18-23, using the Infrared Sideport Imager (ISPI) on the 4m Blanco Telescope at CTIO. The detector is a 2Kx2K HgCdTe HAWAII-2 array which delivers images with a 0.305$''$ pixel$^{-1}$ scale over a field of view (FOV) of $10.5' \times 10.5'$. 

Since the FOV of ISPI is smaller than the $30' \times 30'$ area covered by MUSYC optical images, we divided each field in nine sub-fields (NE, E, SE, S, SW, W, NW, N, C) that we later mosaic as described in Section 4.4. 

The dominance of the sky background and its rapid variability determines the observing strategy in the K band. Dithered short exposures are necessary in order to subtract the background and keep the number of counts in the linear regime of the detector. Individual raw frames are the result of co-adding short exposures. Our science frames have integration times of $4 \times 15$, $5 \times 12$ or $6 \times 10$ seconds, depending on the intensity of the sky background at the moment of the observations. We used a quasi-random dither pattern inside a 45$''$ box.

Dome flats and dark frames were obtained at the beginning and the end of every night. Flats are built by taking the difference between the combined flat frames taken with the lamp on and the combined frames taken with the lamp off in order to subtract the thermal background.

\begin{figure}[ht]
\begin{center}
\plotone{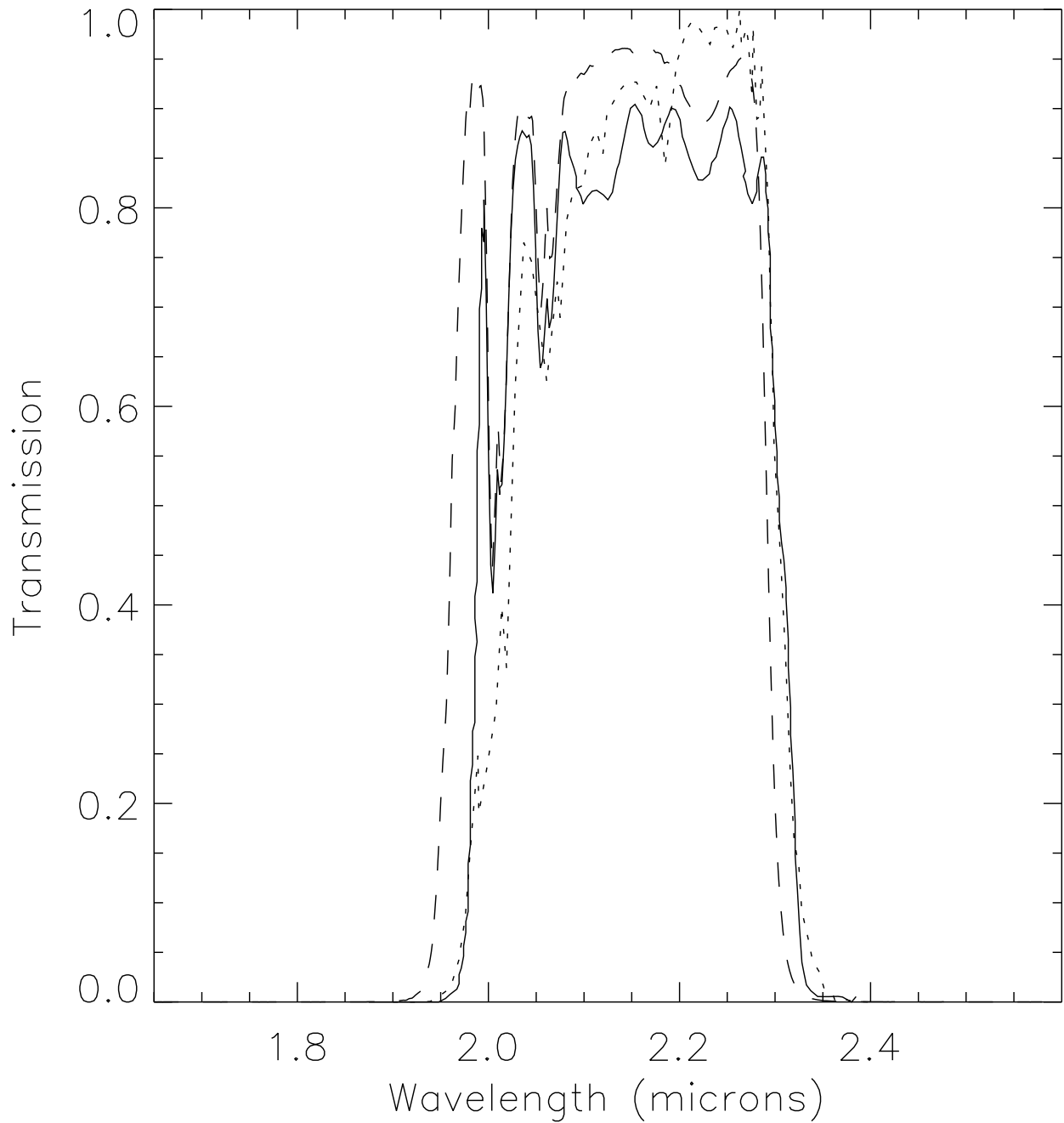}
\end{center}
\caption{Transmission curves of the K' (dashed) and the Ks (solid) filters after multiplying by the detector's quantum efficiency and the atmospheric transmission. The transmission for the Ks 2MASS filter (dotted) is also shown.}
\label{fig-1}
\end{figure}

The K' filter on ISPI was replaced in April 2004 by a new Ks filter. Around 20\% of our observations were completed using the K' filter, and the rest of the survey was completed using the new Ks filter. Figure \ref{fig-1} shows the transmission curves of the two filters after multiplying by the detector quantum efficiency and the atmospheric transmission at CTIO. The difference in effective wavelength for the two filters is $\sim 30$\AA , negligible compared to the filters width. \cite{quadri07b} calculated a conversion between both filters of 0.02 mag, smaller than our typical photometric uncertainties. Therefore in the following discussion we do not distinguish between the two filters and we simply refer to them as the K filter.

\section{Data Reduction and Image Properties}

\subsection{Image Processing}

Data reduction was performed using standard IRAF\footnote{IRAF is distributed by the National Optical Astronomy Observatories, which are operated by the Association of Universities for Research in Astronomy,Inc., under cooperative agreement of the National Science Foundation.} routines, modified IRAF routines and custom IDL tasks. Our reduction scheme is based on the methods described by \cite{labbe03} and \cite{quadri07b}. The external IRAF package XDIMSUM\footnote{XDIMSUM is the Experimental Deep Infrared Mosaicing Software package developed by L. Davis, and it is available at http://iraf.noao.edu/iraf/ftp/extern/xdimsum} was used throughout most of the steps of the reduction.

We subtracted dark frames of the same exposure time and number of co-adds from our science images. Flat frames were constructed for each night as described above, but flat fielding was not performed until after background subtraction. This approach is adopted because the multiplicative nature of the flat fielding process scales up the contribution to the noise coming from the uncertainty in the flat frame by the number of counts in the science image \citep{joyce92}. This can be shown by deriving the uncertainty in a given pixel after background subtraction and flat fielding. We consider the two cases where (1) background subtraction is performed before flat fielding and (2) when flat fielding is performed before background subtraction.

Consider a pixel $p$ which has a background flux $b$ and object flux $o$ in the image under construction, but only background flux in the other $N$ frames used to estimate the background. In case 1, the value of the pixel after background subtraction is

\begin{equation}
p=b+o-<\!b\!>
\end{equation}

and the uncertainty is 

\begin{equation}
\sigma_p^2=\sigma_b^2+\sigma_o^2+\frac{\sigma_b^2}{N}
\end{equation}

now, we divide by the flat and we get

\begin{equation}
p=\frac{b+o-<\!b\!>}{f}
\end{equation}

and the final uncertainty in case 1 becomes

\begin{equation}
\sigma_{p,1}^2=\frac{\sigma_b^2+\sigma_o^2+\frac{\sigma_b^2}{N}}{f^2}+\frac{(b+o)^2\sigma_f^2}{f^4}-\frac{2(b+o)<\!b\!>\sigma_f^2}{f^4}+\frac{<\!b\!>^2\sigma_f^2}{f^4}
\end{equation}

Now, in case 2, we first divide by the flat obtaining

\begin{equation}
p=\frac{b+o}{f}
\end{equation}

with an uncertainty

\begin{equation}
\sigma_p^2=\frac{\sigma_b^2+\sigma_o^2}{f^2}+\frac{(b+o)^2\sigma_f^2}{f^4}
\end{equation}

and then subtract the background obtaining

\begin{equation}
p=\frac{(b+o)}{f}-\frac{<\!b\!>}{f}
\end{equation}

with a final uncertainty for case 2 of

\begin{equation}
\sigma_{p,2}^2=\frac{\sigma_b^2+\sigma_o^2}{f^2}+\frac{(b+o)^2\sigma_f^2}{f^4}+\frac{\sigma_b^2}{Nf^2}+\frac{<\!b\!>^2\sigma_f^2}{f^4}
\end{equation}

It can be seen that the noise in case 2 is higher than in case 1 by an amount $2(b+o)<\!b\!>\sigma_f^2/f^4$. Since the typical background level in our images is $b\sim30000$ counts, and after sky subtraction is $\sim0$ counts, it is clear that the contribution from the uncertainty in the flat frame will be strongly suppressed by flat fielding after sky subtraction. In the above derivation we have neglected the contributions of dark current and read noise.

Background subtraction was performed using a two-phase process that avoids overestimating the background around bright objects. A background image is created for every science frame by combining eight dithered frames adjacent in time. The combination is an average excluding the lowest and highest value at every pixel position. These background frames are subtracted from the science images to create preliminary background subtracted images. The positions of six stars are used to calculate the relative offsets between dithered background subtracted frames and the images are then combined after registering them to a common reference using fractional pixel shifts.

Objects are then detected in the combined image using a thresholding algorithm and an object mask is created. The mask is deregistered to the reference of each original dithered science frame and the background subtraction process is repeated. In this second phase, the background image for each science frame is created as above but masking the detected objects in the time adjacent frames before combining them. This background frame is then subtracted from the science image. 

A second set of stars in these final sky subtracted images is used to measure the FWHM and calculate the relative flux scale between the images. The central $300 \times 300$ pixel region of each frame is used to calculate the pixel-to-pixel rms of the background. These values are used to compute weights used to optimize the signal-to-noise in the seeing disk during the combination of the background subtracted images. The weights are given by,

\begin{equation}
w_i = \frac{1}{(flux\_scale_i\cdot rms_i \cdot FW\!H\!M_i)^2}
\end{equation}

A bad pixel mask (BPM) is created by flagging all $4 \sigma$ deviant pixels in the flat. XDIMSUM uses a cosmic ray detection algorithm and combines this information with the BPM to create a rejection mask for each science frame. Satellite trails found during the visual inspection of the images are added by hand to the corresponding rejection mask. The ISPI detector is not able to properly flush extremely bright pixels during readout, retaining memory of previous exposures in the cores of bright objects. To avoid residuals from previous exposures we construct a previous frame bright object mask for each science image. 

All background subtracted frames were visually inspected in order to reject bad quality images. There are two main reasons for rejection. The first one is catastrophic background subtraction on images due to extreme variations of the background. The second cause of rejection is a poor shape of the point spread function (PSF) due to focusing and tracking problems. Around 10\% of our images were rejected because of these problems. All the accepted background subtracted images are then flat fielded dividing by the normalized dome flat. 

Flat fielded images usually show residuals caused by variability of the bias structure of ISPI's detector. We remove these features by fitting a low order polynomial surface to each quadrant of the images after masking the objects and subtracting this surface from the images. Finally all the images are registered and combined using the above weights, rejecting bad pixels, cosmic rays and residuals from previous frames. An exposure time map is also generated. Details about the K-band images obtained for each sub-field in 1030 and 1255 are given in Table \ref{tbl-2}.

\subsection{Astrometric Projection, PSF Matching and Photometric Calibration}

In order to measure accurate colors, we need to perform aperture photometry in different bands over the same physical region projected on the sky. The MUSYC optical images have been re-sampled to a tangent plane projection with a uniform pixel scale and show an rms astrometric error of 0.2$''$ \citep{gawiser06a}. We re-projected the K-band images so they share the same projection, image size and world coordinate system of the optical images. 

We used the standard IRAF tasks GEOMAP and GEOTRAN to compute and apply a transformation between the logical coordinate system of the ISPI images and the logical system of the reference MUSYC BVR image. For each sub-field image we used the positions of $\sim 100$ stars detected in both the K-band and the BVR image to fit the coefficients of the 2D 6th order Chebyshev polynomials used for the transformation. The coordinates of the stars in the transformed K-band images show an average rms of 0.05$''$ with respect to the coordinates in the BVR image, so our astrometric error is dominated by the uncertainty in the optical images. After re-sampling the pixel scale changes from 0.305$''$ pixel$^{-1}$ to 0.267$''$ pixel$^{-1}$.

The measurement of accurate colors requires matching the PSF of the optical and K-band images. We built an empirical PSF for each K-band sub-field image by registering and combining the images of $\sim$40 stars. We used the same method with $\sim 120$ stars to build empirical PSF for the optical images in each band for the two fields. Tables \ref{tbl-2} and \ref{tbl-3} present the FWHM of the empirical PSF of the K-band and optical images respectively. Since the PSF shows broader wings than a Gaussian profile of the same FWHM, convolution with a purely Gaussian kernel is not able to match both the core and the wings of the PSF at the same time. This implies the need to use a non-Gaussian kernel to obtain accurate PSF matching. We matched the PSF of all the images to a Moffat profile with a FWHM of 1.2$''$ for 1030 and 1.3$''$ for 1255. These target PSF were chosen to be slightly broader (both in the core and the wings) than the PSF of the image with the largest FWHM in the field.

We used the IRAF task LUCY to compute the appropriate convolution kernels for the PSF matching using the Lucy-Richardson iterative algorithm \citep{richardson72, lucy74}. LUCY deconvolves the target PSF into the input PSF and a convolution kernel. The kernels are then normalized and convolved with the images to match their PSF to the target Moffat profile. In order to quantify the quality of the method we constructed growth curves for all the stars used to build the empirical PSF and median combined them to create a master growth curve for each PSF matched image. After convolution the master growth curve matches the Moffat profile growth curve to better than 1\% at all radii for all the images. 

We flux calibrated our images using the 2MASS survey \citep{skrutskie06}. As can be seen in Figure \ref{fig-1} the 2MASS K-band is very similar to the ISPI K-band used in this work. We have computed and applied a correction of 0.03 magnitudes to the zero-points to account for the difference in the filters, which is of the order of our photometric errors. In each of our K-band sub-fields we performed photometry in a 14$''$ diameter aperture of an average of 20 stars with 2MASS counterparts and not affected by saturation, non-linearity or close neighbors in the MUSYC imaging. We calculated the zero-points of each sub-field image by comparing our 14$''$ aperture magnitudes to the aperture-corrected K-band magnitudes of the same stars in the 2MASS point-source catalog (which were measured in 4$''$ apertures and corrected using a curve-of-growth measured out to 14$''$). Uncertainties in the zero-points were computed based on the deviation of the zero-point values obtained for single stars and are typically $\sim 0.02$ mag. Table \ref{tbl-2} presents the zero-point values for each sub-field in 1030 and 1255. All sub-fields were scaled to the zero-point with the lowest uncertainty in the field so the zero-points of our final K-band images are 22.03 for 1030 and 22.38 for the 1255 field.

\subsection{Correlated Noise Properties}

Derivations of photometric uncertainties typically assume that the background noise properties are characterized by Poisson statistics so that the background noise in a given aperture is simply given by $\sqrt{n_{pix}}\sigma_{pix}$, with $\sigma_{pix}$ the pixel-to-pixel rms and $n_{pix}$ the number of pixels in the aperture. This assumption is only valid when the pixels are uncorrelated and can be treated as independent experiments. Sub-pixel shifts used to register the images, astrometric re-projection and PSF matching introduce correlations between pixels. Previous work has shown that assuming Poisson statistics in the presence of correlated pixels significantly underestimates the uncertainties in the background \citep{labbe03,gawiser06a,quadri07b}. Proper characterization of the noise properties of our images is of great importance since the quality of $\chi^2$ fits during photometric redshift estimations relies on the photometric uncertainties.

\begin{figure}[ht]
\begin{center}
\plotone{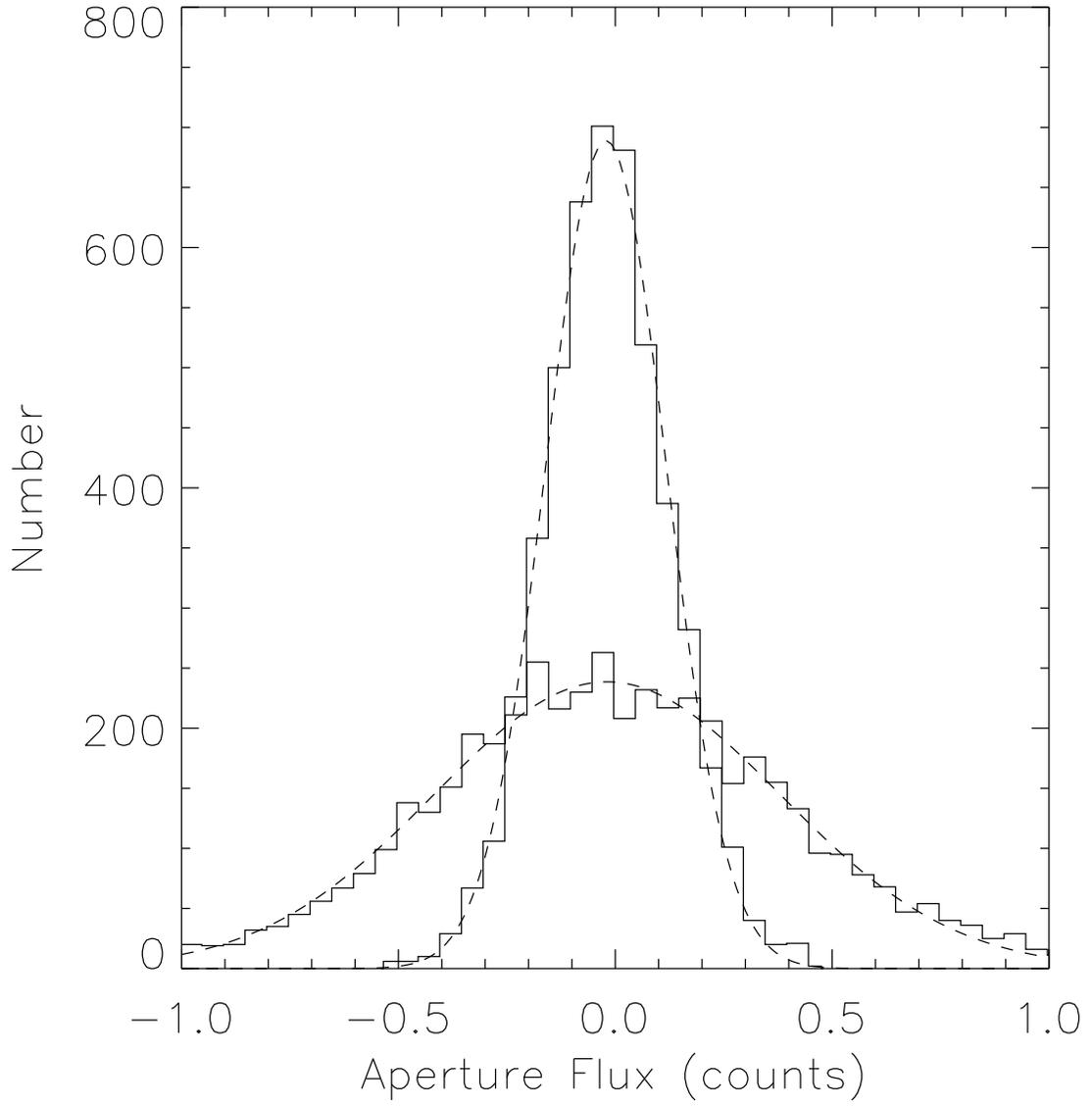}
\end{center}
\caption{Histogram of enclosed fluxes measured on $\sim$5000 empty apertures
  on the 1030 NE sub-field, for two aperture sizes of 0.6$''$ and
  1.1$''$ in diameter. Best fit Gaussians are also shown.}
\label{fig-2}
\end{figure}

We use the ``empty aperture'' method \citep{labbe03} to empirically determine the noise properties of our NIR sub-fields and optical images. Briefly, we randomly place a large number (5000) of apertures of area $n_{pix}$ on our images, reject all apertures falling on sources, and measure the flux in the remaining ones. A histogram of the measured fluxes is constructed and we perform a Gaussian fit to it. The rms of the background in an aperture of area $n_{pix}$ is given by the $\sigma$ of the Gaussian fit. We repeat the process six times for each image and take the average $\sigma$. The method is illustrated by Figure \ref{fig-2}, which shows the histogram of enclosed fluxes for apertures of two different sizes. It is clear that the background noise is well described by a Gaussian distribution. Tables \ref{tbl-2} and \ref{tbl-3} give the $5 \sigma$ point-source limiting depth of our images for an aperture size of $1.4\times$FWHM (see Section 5.2).

\begin{figure}[ht]
\begin{center}
\plotone{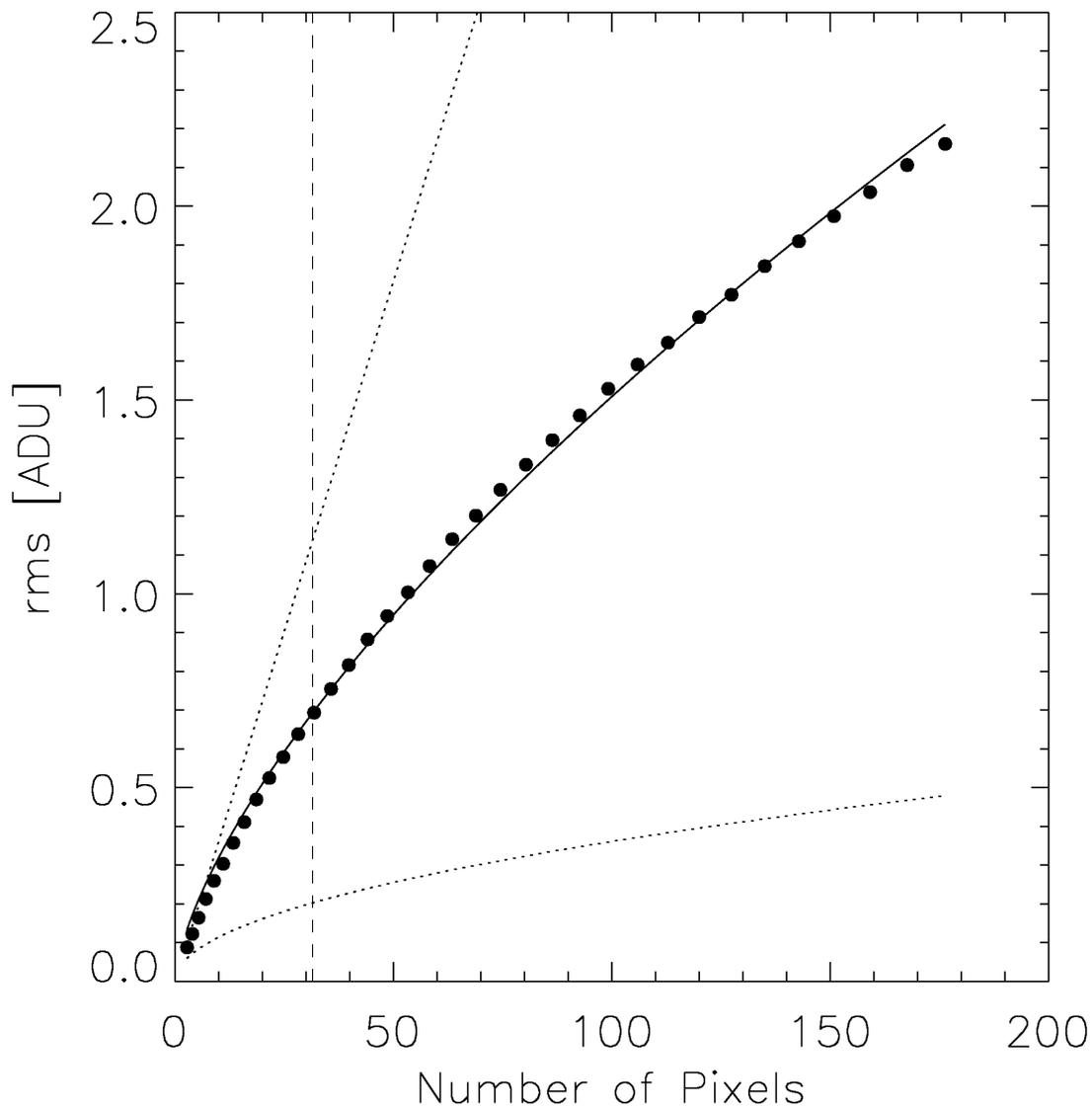}
\end{center}
\caption{Measured background rms (dots) as a function of the number of pixels in the aperture, and best fit (solid line) for the K-band mosaic of the 1030 field. The dotted lines show the formal Poisson scaling dependence with $\sigma_n \alpha \sqrt{n_{pix}}$ (bottom) and a limiting case with $\sigma_n \alpha n_{pix}$ (top). The vertical dashed line corresponds to the aperture used for color photometry (see Section 5.2)}
\label{fig-3}
\end{figure}

We used the same method to measure the dependence of noise with aperture size on the final K-band mosaics (built as explained in Section 4.5). We measured the rms on a series of increasing apertures with diameters between 0.5$''$ and 4.0$''$ (see Figure \ref{fig-3}) and fitted the dependence using the following parametrization,

\begin{equation}
\sigma_{n}=\alpha \sigma_1 n_{pix}^\beta
\end{equation}

where $\sigma_{1}$ is the pixel-to-pixel rms, $n_{pix}$ is the number of pixels in the aperture and $\alpha$, $\beta$ are free parameters. In the Poissonian case we expect $\alpha=1$ and $\beta=0.5$. We obtain $\alpha=1.84$ in both fields and $\beta=0.67$ in 1030 and 0.71 in 1255 respectively. Our result is consistent with previous works showing that assuming Poisson noise in the presence of correlated pixels dramatically underestimates photometric uncertainties.

We built an rms map for each K-band sub-field by taking the inverse square root of the normalized exposure map and scaling it to the rms measured in the $1.4\times$FWHM diameter aperture (the choice of this aperture is discussed in Section 5.2). We properly combined these maps in order to build a mosaic rms map for the whole field and use it to obtain the photometric uncertainty in a given aperture when performing photometry on our sources.

\subsection{Image Mosaicing}

After the astrometric re-projection of the K-band sub-fields we obtained images with the same size and physical coordinate system as the MUSYC optical images, so we just needed to combine the images of each sub-field in a convenient way in order to create a final K-band mosaic of each field. No registration of the frames was necessary. We calculated a weight for each pointing given by

\begin{equation}
w_j=\frac{1}{(flux\_scale_j\cdot rms_j)^2}
\end{equation}

\begin{figure}[ht]
\begin{center}
\plotone{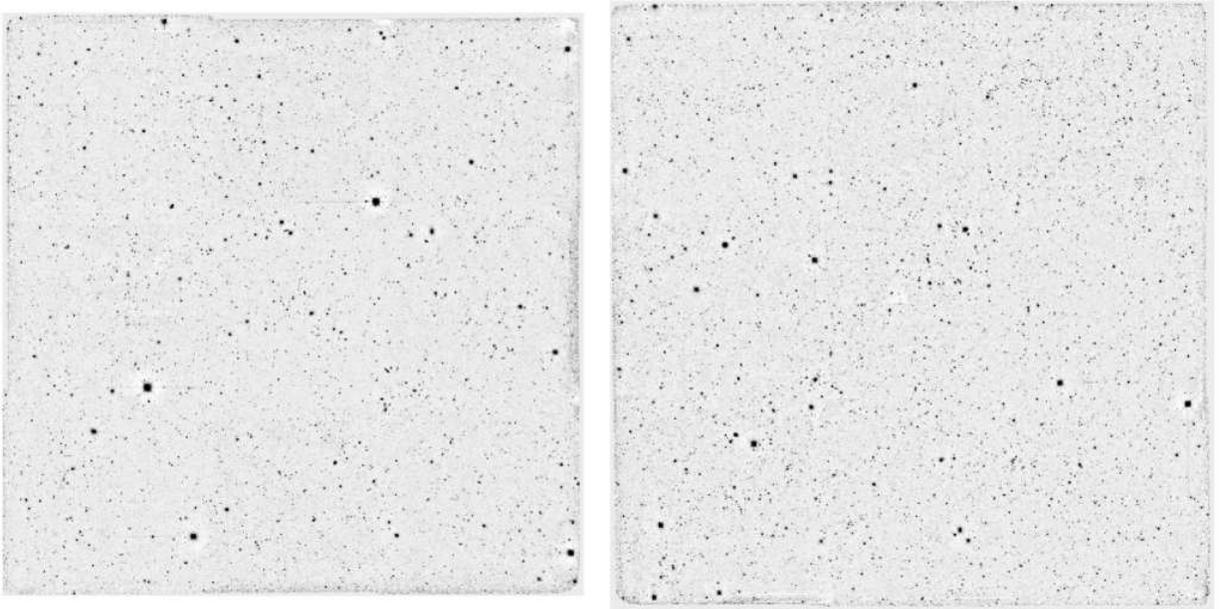}
\caption{Final K-band $\sim30'\times30'$ mosaics of the 1030 (left) and 1255 (right) fields}
\label{fig-4}
\end{center}
\end{figure}

The difference of this weight with the one used in Section 4.1, is that it does not take into account the FWHM since the sub-fields have been PSF matched, and the rms is no longer the pixel-to-pixel rms, but it corresponds to the rms of the background in a $1.4\times$FWHM diameter aperture estimated as described above. The flux scaling factors correspond to the ones used to take all the sub-fields to a common zero-point (see Section 4.3). The normalized exposure map of each sub-field is scaled to the corresponding weight in order to create a final weighting map. Finally, a weighted average of the nine sub-field images is taken at every pixel position in order to create the K-band mosaic. Figure \ref{fig-4} shows the final K-band mosaic of 1030 and 1255.

\section{Source Detection, Photometry and K-selected Catalogs}

\subsection{Detection}

Source detection was performed on the rms normalized K-band mosaic of each field using SExtractor v2.5.0 \citep{bertin96}. The detection image was built by dividing each K-band mosaic by its normalized rms map and then masking the noisy edges of the image (product of the dithering) as well as the regions around bright saturated stars. After masking, the K-band images cover an area of 830 arcmin$^2$ in 1030 and 828 arcmin$^2$ in 1255.

In order to find an optimal set of detection parameters (DETECT\_THRSH, the detection threshold in pixel-to-pixel rms units, and DETECT\_MINAREA, the number of contiguous pixels that must meet the threshold), we estimated the number of spurious sources in our catalogs by running SExtractor with the same set of parameters on the negative of the detection image, which is a robust estimation given the gaussian noise properties of the background. We looked for a set of parameters that maximizes the number of detected sources, while keeping the fraction of spurious sources at $K<20$ under 1\%. We found that filtering the images with a PSF sized kernel lowered the number of detected sources (keeping our spurious fraction constrained) so we did not apply any kind of filtering during the detection. The optimal detection parameters found were DETECT\_MINAREA=1 and DETECT\_THRSH=3.8 (3.6) for 1030 (1255).

\subsection{Photometry and Color Estimation}

\begin{figure}[ht]
\begin{center}
\plotone{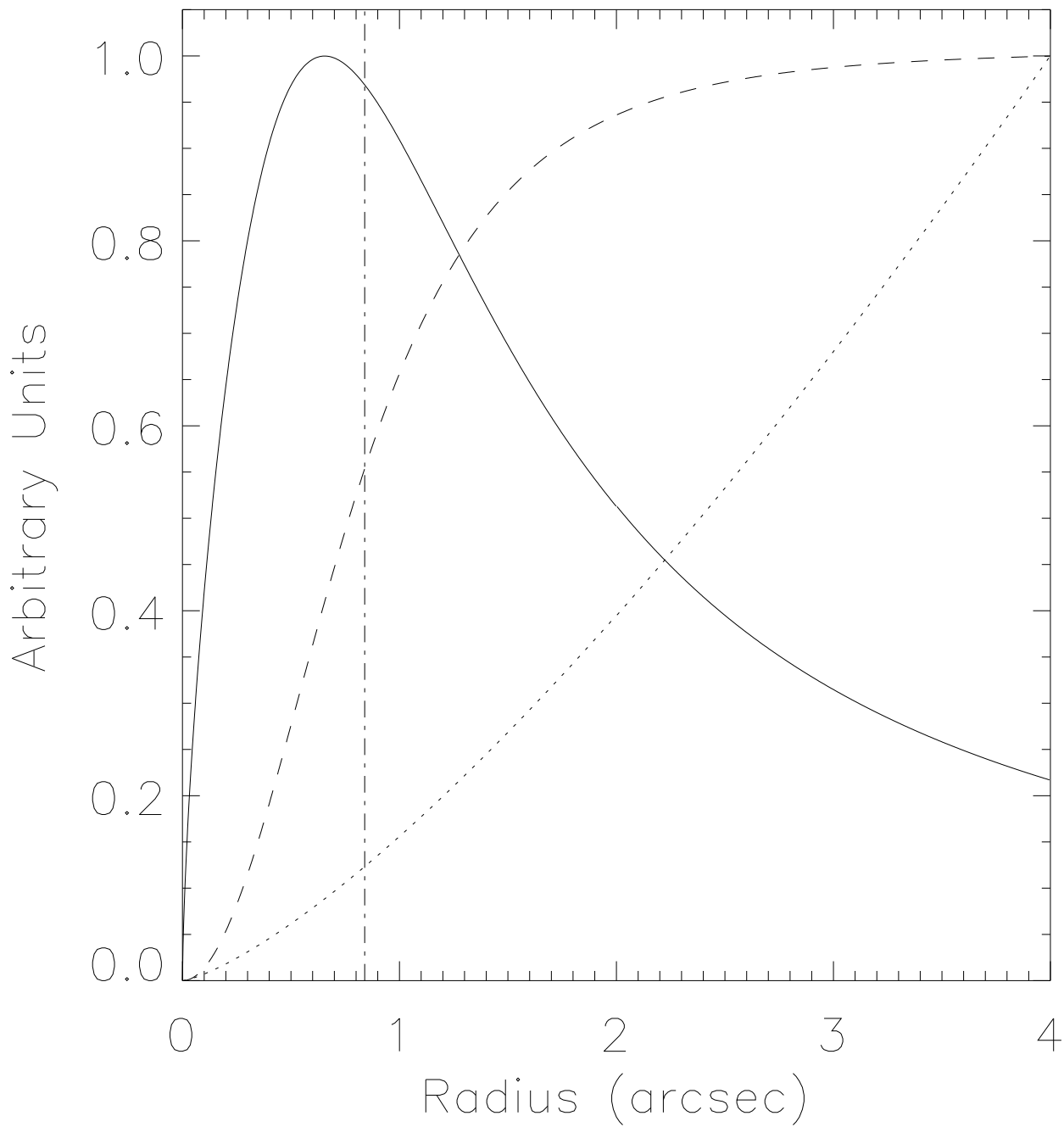}
\end{center}
\caption{The solid line shows the Signal-to-Noise for a point-like source as a function of aperture size for the 1030 field. The dashed curve corresponds to the enclosed flux at a given radius for the field PSF (after PSF matching). The dotted line shows the noise as a function of aperture size presented in Figure \ref{fig-3}. The vertical dashed-dotted line marks the radius of the $1.4\times$FWHM color aperture used for the photometry.}
\label{fig-5}
\end{figure}

Figure \ref{fig-5} shows the normalized Signal-to-Noise (S/N) ratio as a function of aperture size for a point-like source in the 1030 field. Following \cite{gawiser06a} the S/N ratio is built using the growth curve of the Moffat profile to which the PSF was matched and the noise dependence with aperture size ($\sigma_n(n_{pix})$). For 1030 and 1255, the S/N ratio is maximized by an aperture of $\sim 1.1 \times$FWHM. The choice of the aperture used to calculate the colors of our sources is a delicate issue. Choosing a small aperture of $1.1 \times$FWHM maximizes the S/N, but makes the photometry highly dependent on the errors in the PSF matching and astrometric re-projection of the K-band images. We decided to take a more conservative approach and use a ``color aperture'' of $1.4\times$FWHM (given by the half-light radius of the PSF) for which the S/N ratio is still above 95\% of its maximum.

We performed photometry using SExtractor in dual image mode. In this mode, the detection is performed in the K-band detection image and photometric measurements are obtained on the images in all bands. We measured the total K-band flux of our sources using a Kron auto-scaling elliptical aperture (SExtractor's FLUX\_AUTO). It has been shown that Kron apertures miss a small fraction ($<$10\%) of the total flux of extended objects \citep{bertin96}. We accounted for this loss by applying an aperture correction computed using the growth curve of the PSF and correcting by the fraction of the flux falling outside the characteristic radius of the Kron ellipse given by $r_{kron}=\sqrt{ab}$ (with a, b the major and minor semi-axis). Corrections are smaller than 0.1 mag for K$<$20 sources.

Finally colors were estimated using the flux measured in a fixed $1.4\times$FWHM diameter circular ``color aperture'' in all bands. Colors, together with the total K-band flux allow the determination of total fluxes in all bands.

\subsection{K-selected Catalogs}

K-band selected catalogs are publicly available at {\it http://www.astro.yale.edu/musyc/}. All fluxes and uncertainties are given in units of 0.363 $\mu$Jy so the zero-point for AB magnitudes is 25 mag. Photometric uncertainties given in the catalogs correspond to the background rms in the corresponding aperture ($\sigma_{bkg}$). Total photometric uncertainties should be calculated as

\begin{equation}
\sigma_{phot}=\sqrt{(\sigma_{bkg})^2+FLUX }
\end{equation}

 Version 1.0 of the catalogs has the following format:\\
{\it Column 1:}     SExtractor ID number\\
{\it Columns 2-3:}  x and y centroid (pixels)\\
{\it Columns 4-5:}  $\alpha$ and $\delta$ (J2000.0, decimal degrees)\\
{\it Column 6:}     Internal field code: 1=1030 2=1255\\
{\it Columns 7-20:} Flux density and error in color aperture (UBVRIzK)\\
{\it Column 21-22:} K-band aperture-corrected total flux density and error\\
{\it Column 23-29:} Exposure time weight, normalized to maximum for the field (UBVRIzK)\\
{\it Column 30:}    Color aperture diameter (arcsec)\\
{\it Column 31:}    AUTO aperture diameter $2\sqrt{ab}$ (arcsec)\\
{\it Column 32:}    Semi-major axis of AUTO aperture (pixels)\\
{\it Column 33:}    Semi-minor axis of AUTO aperture (pixels)\\
{\it Column 34:}    AUTO aperture Position Angle (decimal degrees)\\
{\it Column 35:}    AUTO aperture correction factor\\
{\it Column 36:}    SExtractor star/galaxy class\\
{\it Column 37:}    Object was originally blended with another\\
{\it Column 38:}    Object's neighbors may significantly bias AUTO photometry\\
{\it Column 39:}    SExtractor flag for the K-band

\section{Star-Galaxy Classification}

We separated stars from galaxies in our catalogs using the two color criteria introduced by \cite{daddi04} in which stars populate a region in the $(B-z)_{AB}$ versus $(z-K)_{AB}$ plane (hereafter: BzK plane) given by $(z-K)_{AB}<0.3(B-z)_{AB}-0.5$. This method was calibrated using the spectroscopically complete K20 survey \citep{cimatti02c} and has been shown to be extremely robust, showing a very high efficiency ($>$90\%) at selecting stars and very little contamination \citep{daddi04}. Since the BzK method was developed by Daddi using a slighltly different set of filters than the ones used in this work (the biggest difference is the use of a Bessel B-band as opposed to the Johnson B-band used here) we used the (B-z) and (z-K) colors of 35 stars in the K20 GOODS-South field given by \cite{daddi04} and compared them to the colors of the same stars obtained from our B, z and K-band photometry in the ECDF-S field \citep{taylor07}. We apply the computed offsets of  -0.04 mag in $(z-K)$ and 0.56 mag in $(B-z)$ to our colors before plotting them in the BzK diagram.

\begin{figure}[ht]
\begin{center}
\epsscale{0.8}
\plotone{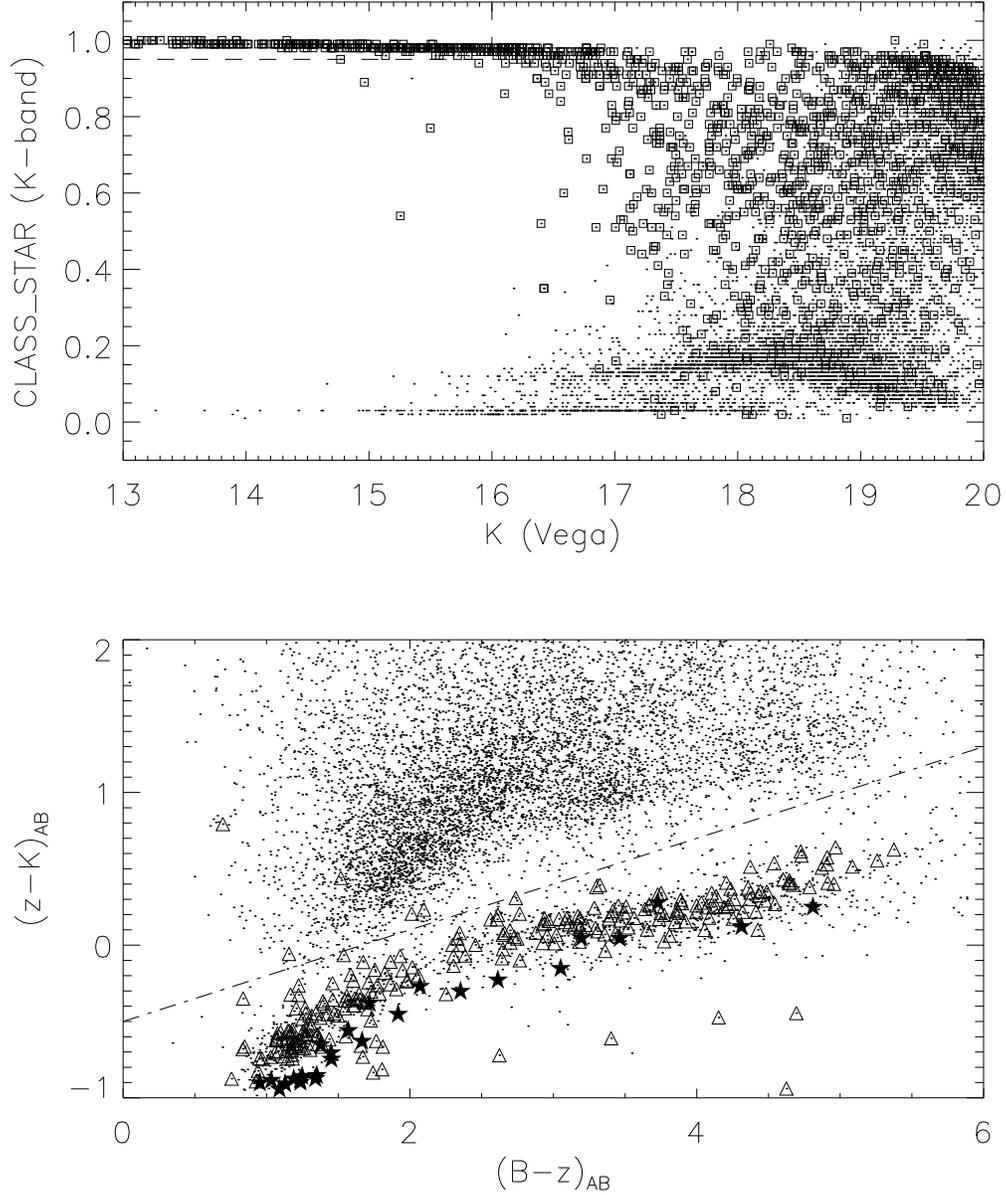}
\caption{Top: K-band magnitude versus SExtractor's CLASS\_STAR parameter for sources brighter than $K=20$ (dots) in the 1255 field. Objects with CLASS\_STAR $\sim$1 are supposed to be point-like. The dashed lines show the $K<16$ and CLASS\_STAR$>$0.95 region where the separation works well. Squares correspond to objects classified as stars using the $(z-K)_{AB}<0.3(B-z)_{AB}-0.5$ criteria. Bottom: BzK diagram for sources brighter than $K=20$ (dots) in the same field. Triangles show sources with $K<16$ and CLASS\_STAR$>$0.95. Stars correspond to stars of different spectral types from \cite{pickles98}. The dashed-dotted line shows the $(z-K)_{AB}=0.3(B-z)_{AB}-0.5$ criteria used to separate stars from galaxies.}
\label{fig-6}
\end{center}
\end{figure}

This method works better on our data than the star-galaxy classification based on the SExtractor CLASS\_STAR parameter. The lower panel of Figure \ref{fig-6} shows the BzK diagram for sources brighter than $K=20$ in the 1255 field. We have overploted stars of different spectral types (from O5 to M6) taken from \cite{pickles98}, all of which fall in the selection region. From the lower panel in Figure \ref{fig-6} we can conclude that the CLASS\_STAR parameter works correctly on our data up to $K\sim16$. At fainter magnitudes the method loses its ability to separate stars from galaxies. 

After star-galaxy separation we detected 1344 stars and 8015 galaxies in 1030 and 1547 stars and 8115 galaxies in 1255 to a magnitude limit of $K=20$. The number of stars detected is in rough agreement with models of stellar population synthesis in the Galaxy \citep{robin03} which predict a number of $\sim 1100$ stars brighter than $K=20$ for the areas and galactic coordinates of our fields.

\section{Number Counts}

\begin{figure}[ht]
\begin{center}
\plotone{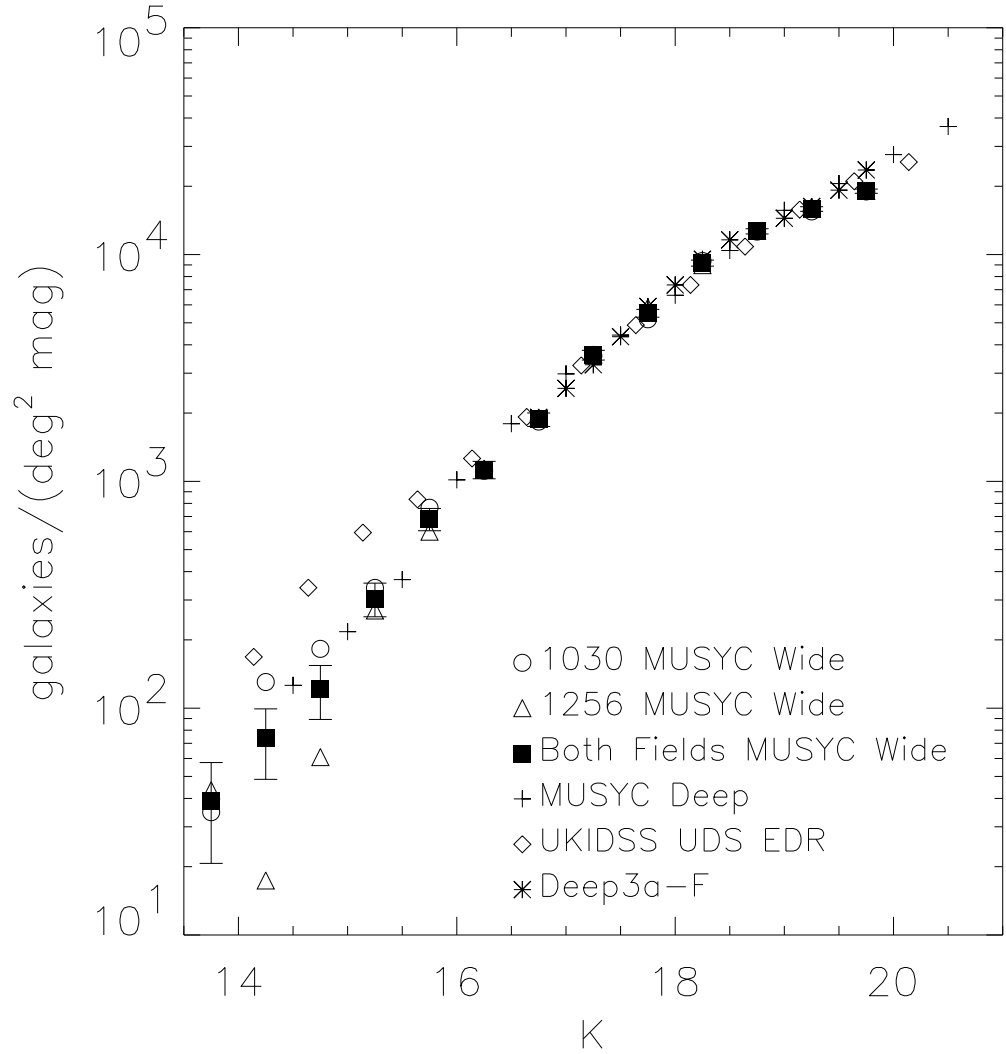}
\caption{Differential K-band number counts for galaxies in 1030 and 1255, together with data from previous works by \cite{kong06,quadri07b,lane07}}
\label{fig-7}
\end{center}
\end{figure}

We computed differential K-band number counts in 0.5 magnitude bins for all galaxies in both fields separately. The results are presented in Figure \ref{fig-7} together with data from the UKIRT Infrared Deep Sky Survey (UKIDSS) Ultra Deep Survey Early Data Release (UDS EDR) from \cite{lane07}, the Deep3a-F field from \cite{kong06} and the MUSYC Deep NIR Survey \citep{quadri07b}. Error bars assume Poisson statistics and underestimate the real errors because cosmic variance is not taken into account. The 1030 and 1255 fields show an excellent agreement (within 1$\sigma$) in their number counts in the $16 \le K \le 20$ range. We do not observe the excess in the galaxy number counts observed by \cite{quadri07b} in the 1030 field, showing how larger areas (the MUSYC Wide fields are nine times larger than the MUSYC Deep fields) help to overcome the problem of cosmic variance. We do, however, observe an excess of passive galaxies at $z \sim 2$ in this field as reported in Section 9.1. The flux distribution of galaxies in our sample shows an excellent agreement with previous works. From the comparison with the UKIDDS UDS and the MUSYC Deep NIR Survey, both of which are $\sim 1$ mag deeper than our survey, we conclude that our sample of K-selected galaxies $\sim$80\% complete in the faintest magnitude bin ($19.5<K<20$).

\section{The BzK selection technique}

\begin{figure}[ht]
\begin{center}
\plotone{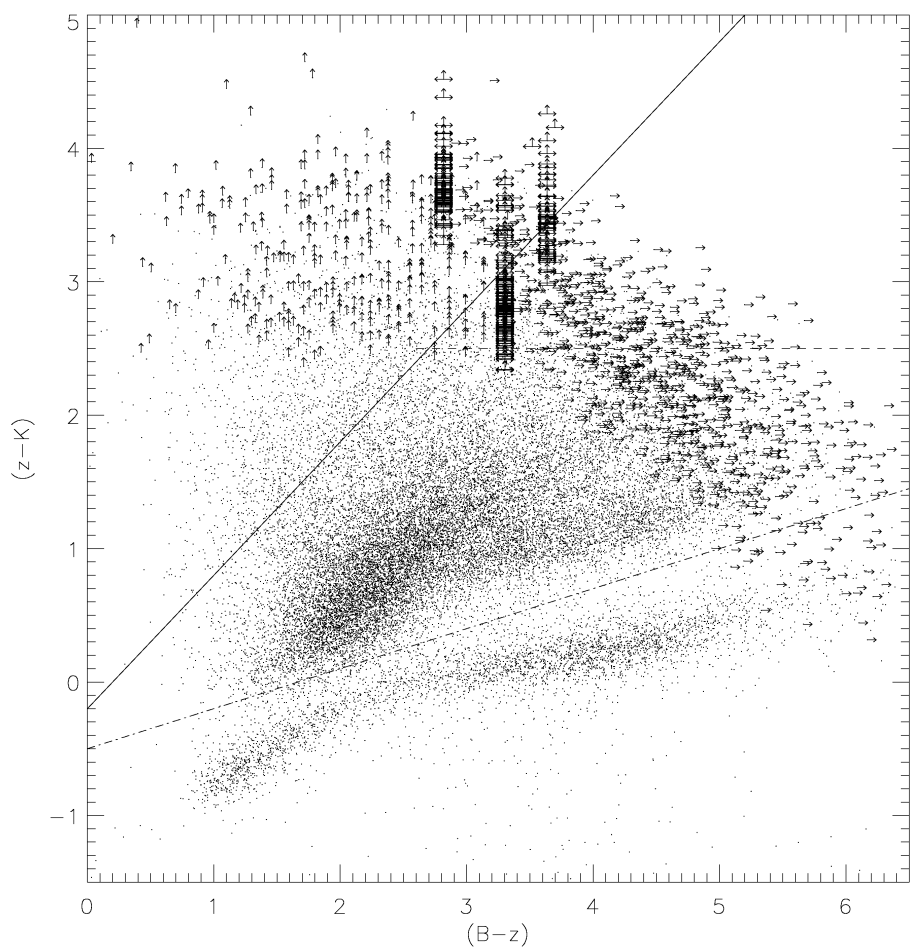}
\caption{BzK plane. Dots show the corrected colors of all $K<20$ sources in the three fields of the MUSYC Wide NIR Survey. The diagonal solid line delimits the sBzK selection region ($BzK \geq -0.2$). The horizontal dashed line defines the pBzK selection region ($(z-K)_{AB}>2.5$). The diagonal dashed-dotted line shows the star-galaxy separation criterion ($(z-K)_{AB}<0.3(B-z)_{AB}-0.5$).}
\label{fig-8}
\end{center}
\end{figure}

\cite{daddi04} introduced a two-color selection criteria based on B, z and K-band photometry to select star-forming and passively evolving galaxies in the $1.4 \leq z \leq 2.5$ range. This method is known as the BzK selection technique, and presents a series of advantages like being almost unbiased against highly reddened galaxies and being able to separate star-forming from passive galaxies. Daddi used the 94\% spectroscopically complete sample of 311 $K<20$ galaxies from the K20 survey which covered 32 arcmin$^2$ in the GOODS-South field \citep{cimatti02c} to calibrate the method. Star-forming galaxies at $z>1.4$ (sBzK) occupy a well defined region in the BzK plane to the left of the solid line shown in Figure \ref{fig-8} defined by $BzK\equiv(z-K)_{AB}-(B-z)_{AB} \geq -0.2$. Old and passively evolving galaxies at $z>1.4$ (pBzK) are located on the upper right corner of the BzK plane, to the right of the solid line and above the dashed line in Figure \ref{fig-8} in a region defined by $BzK < -0.2 \cap  (z-K)_{AB} > 2.5$. As stated in Section 6, stars are clearly separated from galaxies (especially from those with $z>1.4$), and are confined to the region under the dashed-dotted line in Figure \ref{fig-8}, defined by $(z-K)_{AB}<0.3(B-z)_{AB}-0.5$. Daddi showed that the method is highly efficient at selecting z$\sim$2 galaxies and presents little contamination from low-z interlopers. The reddening vector in the BzK plane is approximately parallel to the sBzK selection criteria, which ensures that the method is not biased against heavily reddened dusty galaxies.

\section{Results}

\subsection{BzK Galaxies in the MUSYC Wide NIR Survey}

We have used our data in the 1030 and 1255 fields together with the MUSYC Wide data in the ECDF-S to select BzK galaxies over a total area of 0.71 deg$^2$ to a limiting depth of $K<20$. A K-band selected catalog was produced for the ECDF-S using very similar techniques to the ones used here, and will be presented by \cite{taylor07}. By combining the three MUSYC Wide NIR Survey fields, this is the largest area ever surveyed for BzK galaxies. Since \cite{daddi04} used a slightly different set of filters than the ones used here we found the need to apply a correction to our color photometry in order to be able to use the BzK selection criteria consistently. We have used the $(B-z)$ and $(z-K)$ colors of stars in the K20/GOODS field (which is completely covered by our ECDF-S imaging) provided by \cite{daddi04} to compute offsets that we then applied to our measured colors. After correcting our photometry we applied the BzK criteria to our data to produce a large sample of $z\sim2$ star-forming and passive galaxies. 

Figure \ref{fig-8} presents the positions of all $K<20$ sources in the MUSYC Wide NIR Survey in the BzK plane. 291 objects show no detections to the $1 \sigma$ level in the B and z bands, appearing as lower limits in $(z-K)$ and having an undetermined $(B-z)$ color in Figure \ref{fig-8}. They correspond to 1.0\% of the detected sources. This fraction agrees with our expected 1\% of spurious sources (see Section 5.1) so we assume most of these objects correspond to spurious detections and eliminate them from our analysis. Galaxies falling in the sBzK region and showing lower limits in $(B-z)$ as well as those falling in the pBzK region with lower limits in $(z-K)$ cannot be unambiguously classified. There are 73 of these unclassified objects and for simplicity we leave them outside any further analysis. Including these objects does not affect any of our following results in a significant manner. Out of the 24399 detected galaxies, we have unambiguously selected a sample of 3261 BzK galaxies (2502 sBzK and 759 pBzK) at $z \sim 2$. This is the largest existing sample of BzK galaxies to this depth.

Over the whole surveyed area sBzK galaxies have a sky density of 0.98$\pm$0.06 arcmin$^{-2}$, where we have used the relative deviation between our fields to include cosmic variance when estimating the error. This value is lower than the previous value of 1.2$\pm$0.05 arcmin$^{-2}$ obtained by \cite{kong06} and consistent with the 0.94$\pm$0.17 arcmin$^{-2}$ value for the K20/GOODS sample \citep{daddi04} (both values showing only Poisson errors). For pBzK galaxies we obtain a sky density of 0.3$\pm$0.1 arcmin$^{-2}$ (again including cosmic variance in the error), consistent with the values obtained by Kong (0.38$\pm$0.03 arcmin$^{-2}$) and Daddi (0.22 $\pm$ 0.08 armin$^{-2}$). Due to the considerably larger area surveyed in this work compared to these previous studies, and the fact that both sBzK and pBzK galaxies are strongly clustered as we show below, our values for the sky density of these objects are much less affected by cosmic variance and are therefore more representative of the true sky density of these objects.

Table \ref{tbl-4} shows the number of sBzK and pBzK galaxies in our sample and their sky densities for each of our three fields. The GOODS-S field seems to be underdense in DRG at $z>2.5$ \citep{marchesini07a} and in optically bright AGN at high redshifts \citep{dwelly06}. Furthermore, \cite{vandokkum06} found that the sky density of massive ($M>10^{11}$M$_{\odot}$) K-selected galaxies at $2<z<3$ in the GOODS-S field is a factor of 3 lower than that of the $10'\times10'$ MUSYC Deep 1030 field (the densest field in that work). Here we find a similar trend for passive galaxies in the $1.4<z<2.5$ range, where the sky density of pBzK galaxies in the $30'\times30'$ ECDF-S is a factor of 3 lower than in the $30'\times30'$ 1030. The density of passive $z \sim 2$ galaxies in the ECDF-S field is 60\% of the mean value for the whole survey, while in the 1030 field it is a factor 1.7 higher. Interestingly, star-forming galaxies at $z \sim 2$ show no sign of a significant underdensity.

The large number of objects in our sample allows us to clearly identify the new branch of galaxies in the BzK diagram reported by \cite{lane07}. This feature runs parallel to the stellar sequence in the $2.5<(B-z)<5.0$ range at $(z-K) \sim 1.0$. Lane found it to be consistent with the track of passively evolving early-type galaxies at $z<1.4$. This is the first confirmation of the existence of this new branch since it was reported.

\subsection{Redshift Distribution}

Photometric redshifts rely on the presence of continuum spectral features in the galaxy SED strong enough to show up in broad-band photometry. The most prominent features of this kind are the Lyman break at 912 \AA\ and the 4000 \AA\ break. At $z \sim 1.5$ the 4000 \AA\ break is just red-wards of the z-band and the Lyman break is $\sim$1000\AA\ blue-wards from the U-band. As we move to higher redshifts we have no strong spectral features being sampled by our broad-band photometry (UBVRIzK) until we reach $z \sim 2.6$ where the Lyman break enters the U-band (the 4000 \AA\ break does not get sampled by the K-band until $z \sim 4$). This means that using our data alone we are unable to estimate reliable photometric redshifts for our sample of BzK galaxies since they inhabit a nearly identical redshift range to that in which we cannot sample any spectral continuum features. We would not have this problem if we had J and H-band imaging of our fields.

To overcome this problem we have used the data from the MUSYC Deep NIR Survey, which covers a smaller area to a greater depth and includes JHK near-infrared imaging, as well as the UBVRIz optical coverage. All the data in the Deep Survey was taken using the same instrument as the data presented here, and was reduced using analogous techniques. We have limited the Deep Survey catalog to $K<20$ and used the same color corrections and selection criteria explained above to select a sample of 514 BzK galaxies (365 sBzK and 149 pBzK). This sample should be representative of our BzK galaxy sample since there are no significant differences in the selection biases involved.

Photometric redshifts of the Deep Survey galaxies were calculated with the methods described by \cite{rudnick01} using linear combinations of four \cite{coleman80} empirical templates, two starburst templates from \cite{kinney96} and two solar metallicity, dust-free 10Myr and 1Gyr old single stellar population templates \citep{bruzual&charlot03}. For more details on the photo-z estimations refer to \cite{quadri07b}.

\begin{figure}[ht]
\begin{center}
\plotone{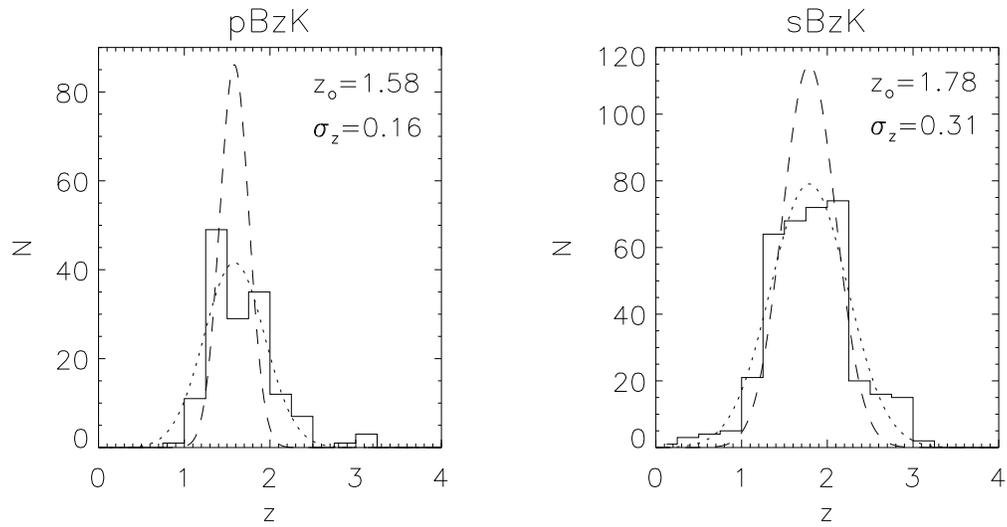}
\caption{Photometric redshift distribution of $K<20$ pBzK and sBzK galaxies in the MUSYC Deep NIR Survey. Dotted lines correspond to the best Gaussian fit to the observed redshift distribution $N_{obs}(z)$. Dashed lines correspond to the distribution after correcting for the broadening due to errors in the photometric redshifts. The parameters reported correspond to the corrected distribution $N_{corr}(z)$}
\label{fig-9}
\end{center}
\end{figure}

Figure \ref{fig-9} shows the redshift distribution $N(z)$ of $K<20$ pBzK and sBzK galaxies in the Deep Survey. Both distributions are fairly Gaussian and cover the expected redshift range for BzK galaxies, roughly $1.4 \lesssim z \lesssim 2.5$. The distribution of sBzK galaxies is clearly broader than the distribution for pBzK. The peak at the $1.25<z<1.5$ bin in the pBzK distribution comes from the 1030 field and is not present in any of the other three Deep Survey fields. As can be seen from Table \ref{tbl-4} the 1030 wide field shows an excess of pBzK galaxies of a factor of 1.7 with respect to the whole survey. This overdensity, together with the presence of the peak in the redshift distribution might indicate the presence of a large scale structure at $z \sim 1.4$ in the 1030 field.

Similar to \cite{hayashi07} we obtain best Gaussian fits of our redshift distributions, but in our case we correct them for the effects introduced by the large errors involved in photometric redshift estimations as explained below and use these to recover the correlation length from the angular correlation function through the inverse Limber projection (see Section 9.5). In that work Hayashi used the spectroscopic redshift distribution of bright ($K<20.1$) sBzK galaxies to compute correlation lengths ($r_0$) for a much deeper sample of sBzK ($K<21.3$) under the assumption that $N(z)$ is not dependent on K magnitude. This latter assumption was proved to be wrong by \cite{quadri07b} using the MUSYC Deep NIR Survey data. By looking at Figure 12 of \cite{quadri07b} it is clear that the redshift distribution of sBzK is severely broadened when going to fainter magnitudes ($K<21$) as the BzK technique loses its power to select galaxies in a well determined and narrow redshift range. This broadening of $N(z)$ for $K>20$ BzK galaxies had also been reported by \cite{reddy05} using spectroscopic redshifts of sBzK galaxies in the GOODS-N field. Assuming a much narrower redshift distribution during the clustering analysis should underestimate the real value of $r_0$, and this might affect the level of clustering measured by Hayashi et al. for their faint sample of sBzK galaxies.

The observed $N(z)$ is strongly affected by errors in the photometric redshifts. In the redshift range of interest the Deep Survey photometric redshifts show a dispersion of $\Delta z/(1+z)=0.12$ when compared to spectroscopic redshifts drawn from the literature \citep{quadri07a}. To a first approximation, under the basic assumption that $N(z)$ is a normal distribution and that photometric redshift errors are of a Gaussian nature, the net effect of the errors is to broaden the intrinsic redshift distribution. The observed distribution in this case is the result of the convolution of the intrinsic distribution and the photometric error distribution. This is a delicate issue because of the assumptions involved in this argument. First, we are assuming that the scatter in the photometric redshift is not redshift dependent, which given the narrow range over which BzK galaxies are selected seems like a reasonable approximation. Also, uncertainties in photometric redshifts are subject to a series of systematics related to the methods used to estimate them (templates used, photometric errors in different bands, aliasing of SED features, etc) which can lead to a non-Gaussian distribution of errors. Nonetheless, unless a very specific and unlikely combination of error distribution and redshift dependence of this distribution occurs, the net effect of photometric uncertainties is to broaden the intrinsic redshift distribution considerably and this effect must be taken into account in order to avoid significant overestimation of the level of clustering of the galaxies in the sample.

The best Gaussian fits to the observed $N(z)$ are centered at $z_0=1.58\pm0.04$ and $z_0=1.78\pm0.03$ and have a width of $\sigma_{z,obs}=0.35\pm0.04$ and $\sigma_{z,obs}=0.45\pm0.03$ for pBzK and sBzK galaxies respectively, and are shown as the dotted lines in Figure \ref{fig-9}. Errors in the parameters come from Monte-Carlo simulations of 1000 renditions of $N(z)$ where the redshift of the galaxies were varied within their uncertainties. We have deconvolved these from our assumed Gaussian distribution of photometric errors of width $\Delta z$, as determined by \cite{quadri07a}, measured at the center of the distribution for each population, to obtain our estimates of the intrinsic redshift distribution $N_{corr}(z)$. The intrinsic distributions are shown as dashed lines in Figure \ref{fig-9}, normalized to the total number of objects in each sample, and have widths of $\sigma_z=0.17\pm 0.06$ and $\sigma_z=0.31\pm0.04$ for pBzK and sBzK galaxies respectively. The latter value is comfortingly consistent with the $0.35$ obtained by \cite{hayashi07} for a sample of 81 sBzK galaxies with spectroscopic redshifts, supporting the validity of the above assumptions.

\subsection{Stellar Masses, Reddening and Star Formation Rates}

Besides calibrating the BzK method, Daddi used \cite{bruzual&charlot03} models to perform SED fitting of the K20/GOODS galaxies with known spectroscopic redshifts to calibrate relations that estimate the stellar mass of BzK galaxies and the reddening of sBzK galaxies based on the B,z and K band photometry alone. He also gave a recipe to estimate the ongoing SFR of sBzK using the B flux (rest-frame UV flux) based on calibrations by \cite{madau98}. For details on the calibration of these methods and the uncertainties involved in them refer to \cite{daddi04}. We have used these relations to calculate stellar masses, reddening (E(B-V)), and SFR of the galaxies in our sample. It is worth noting that, since we do not have redshift information for individual galaxies, these results should be interpreted in a statistical sense, especially the SFR estimates. \cite{daddi04} showed that mass estimates have uncertainties of $\sim$60\% for single objects and E(B-V) estimates show a residual rms of about 0.06 when compared to the values obtained from the multi-band SED fitting of objects with known spectroscopic redshifts. In any case, given the large number of galaxies in our sample and the narrow Gaussian shape of the redshift distributions presented in the previous section, the median stellar masses, E(B-V) and SFR should be highly representative of the properties of typical BzK galaxies at $z\sim2$.

\begin{figure}[ht]
\begin{center}
\epsscale{0.5}
\plotone{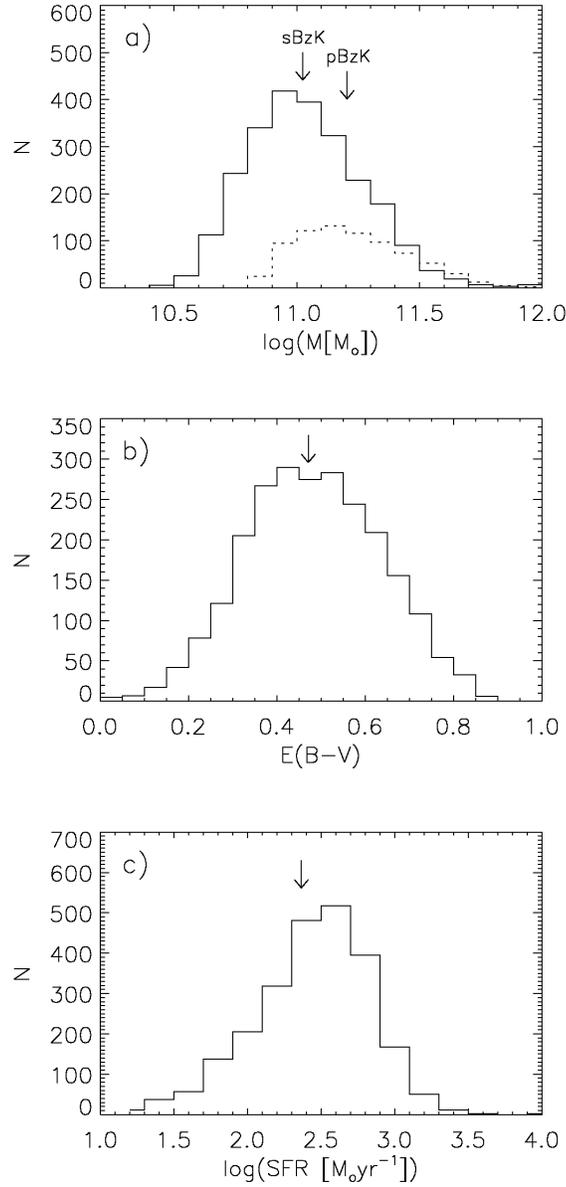}
\caption{a) Distribution of stellar-masses of sBZK (solid) and pBzK (dotted) galaxies. b) Histogram of E(B-V) values for star-forming BzK galaxies in the sample. c) Histogram of SFR calculated using the dereddened B-band flux of sBzK galaxies. Arrows indicate median values of the distributions.}
\label{fig-10}
\end{center}
\end{figure}

The top panel in Figure \ref{fig-10} shows the distribution of stellar masses of sBzK and pBzK galaxies in our sample. It can be seen that sBzK galaxies span a broader range in mass than pBzK, extending down to masses of a few $10^{10}$M$_{\odot}$, while the lowest mass pBzK galaxies in our sample have $\sim10^{11}$M$_{\odot}$. The median mass of sBzK is $\sim1.0\times10^{11}$M$_{\odot}$, while pBzK have a median mass of $\sim 1.6 \times 10^{11}$M$_{\odot}$. These results are in excellent agreement with the results of \cite{kong06} in the Deep3a-F. As noted in that work, the lower limit of the mass distribution of pBzK galaxies corresponds to the mass of a $K=20$ galaxy lying on top of the $(z-K)=2.5$ selection limit, and hence has its origin in the construction of the pBzK sample.

The E(B-V) distribution for sBzK galaxies is shown in the central panel of Figure \ref{fig-10}. Daddi's recipe to estimate reddening uses the $(B-z)$ color as a measure of the UV slope of the SED and assumes a gray and self-similar \cite{calzetti00} extinction law. The median reddening of our sBzK galaxies is E(B-V)=0.47. \cite{adelberger00} present the distribution of extinction at 1600 \AA\ for Lyman-break galaxies at $z\sim3$. Under the Calzetti extinction law assumption it can be seen that practically no galaxies with E(B-V)$>$0.45 are selected by the LBG technique. \cite{reddy05} shows that for spectroscopically confirmed BM/BX galaxies at $z>1$ typical E(B-V) values are lower than 0.3. This means that at least 55\% of the sBzK galaxies in our sample would not be detected by rest-frame UV continuum based selection techniques.

SFR of sBzK were estimated using the B-band flux as the 1500\AA\ rest-frame flux (the B-band effective wavelength corresponds to 1845\AA\ at z=1.4, 1593 \AA\ at z=1.78 and 1265\AA\ at z=2.5) and the calibration by \cite{madau98}. UV fluxes were dereddened using the estimated E(B-V) values for each galaxy. As stated above, the lack of knowledge of the distance modulus of our galaxies means that these SFR estimates are not reliable for individual objects, but the median SFR of the sample should correspond to the typical SFR of sBzK galaxies. The distribution of SFR is presented in the bottom panel of Figure \ref{fig-10}, and it shows a median value of $\simeq$230 M$_{\odot}$yr$^{-1}$. This value is similar but slightly higher than previous estimates by \cite{daddi04} and \cite{kong06}. Given our survey comoving volume of $8.7\times10^6$Mpc$^3$ (roughly $1.4<z<2.5$) we compute a number density of sBzK galaxies of $\sim$3$\times$10$^{-4}$Mpc$^{-3}$, which translates to a SFRD of $\sim$0.07M$_{\odot}$yr$^{-1}$Mpc$^{-3}$. The latest estimates of the SFRD at $z\sim2$ based on rest-frame UV selected samples \citep{reddy07} show a SFRD$\simeq$0.2 M$_{\odot}$yr$^{-1}$Mpc$^{-3}$ at $z=2$. NIR bright sBzK can account for about a 30\% of this total SFRD. Considering that UV selection methods miss aproximately half sBzK galaxies, including bright ($K<20$) star-formig BzK galaxies in the SFRD would increase its value by aproximately a 15\%. 

\subsection{Number Counts of BzK Galaxies}

\begin{figure}[ht]
\begin{center}
\plotone{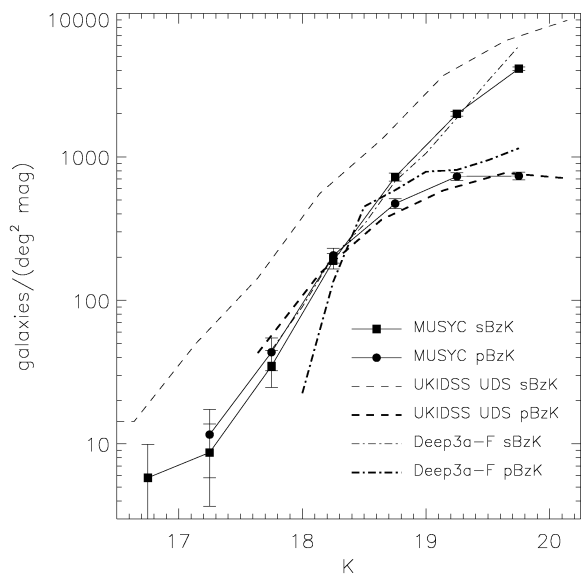}
\caption{Differential K-band number counts of $K<20$ sBzK (squares) and pBzK (circles) in the MUSYC Wide NIR Survey, together with data from previous works by \cite{kong06} and \cite{lane07}.}
\label{fig-11}
\end{center}
\end{figure}

In Figure \ref{fig-11} we present differential K-band number counts for sBzK and pBzK galaxies in our sample, together with previous estimates by \cite{kong06} and \cite{lane07}. The number counts of sBzK galaxies are very steep and the number of objects increases sharply towards higher magnitudes. On the other hand, passive BzK galaxies exhibit a flattening in the number counts towards fainter magnitudes with a knee around $K \sim 18.5$. This behavior has been observed in previous works and is attributed to the fact that because of the narrow redshift distribution of pBzK, the number counts actually probe the luminosity function of these objects. Star-forming BzK galaxies do not show this behavior. This has been usually attributed to sBzK galaxies having a wider redshift distribution than pBzK \citep{kong06, lane07}. Though sBzK have a wider redshift distribution than pBzK in our sample, this difference in the number counts might as well be related to intrinsic differences between the luminosity function of both populations. \cite{marchesini07a} has shown that the faint-end slope of the luminosity function of galaxies with blue rest-frame UV color in the $2<z<2.5$ range is much steeper than for galaxies showing a redder rest-frame UV color. A qualitative comparison of Figure \ref{fig-11} and Figure 5 in \cite{marchesini07a} supports this statement. Further study is necessary to decouple the effects of these differences in the redshift distribution and the luminosity function on the shape of the number counts.

Number counts of pBzK galaxies show an excellent agreement with previous results by \cite{lane07} on the UKIDSS UDS field, and a somewhat worse agreement with the \cite{kong06} sample which can be attributed to the small number of objects in the Deep3a-F, especially at bright magnitudes. On the other hand number counts of sBzK in our sample are in perfect agreement with the sBzK sample on Deep3a-F (a factor of $\sim$3 more numerous than the pBzK sample in the same field). Both Kong's number counts and the ones presented here are systematically lower than the number counts of sBzK galaxies in the UKIDSS UDS sample over the whole magnitude range. The origin of this discrepancy lies in the fact that in this work, as well as in Kong's, BzK colors were corrected in order to take into account the difference in filters used relative to the ones used by \cite{daddi04} to calibrate the BzK method. \cite{lane07} did not apply this correction, and hence their BzK selection criteria was not used consistently with the original calibration of the method. Not applying this correction causes an effective offset of $\sim$0.5 mag towards lower values in the $(B-z)$ colors of their objects (or equivalently an offset of the selection criteria limit towards higher values of $(B-z)$). As can clearly be seen by inspecting Figure \ref{fig-8} this offset will produce a significant excess of objects selected as sBzK but will not affect the pBzK sample in such a radical manner (explaining our agreement with Lane's pBzK number counts). This excess of low-z contaminants in the UKIDSS UDS sBzK sample can account for the discrepancy observed in the number counts.

\subsection{Clustering of BzK Galaxies}

\subsubsection{The Angular 2-Point Correlation Function} 

The angular 2-point correlation function $\omega(\theta)$ is a useful statistic that helps us characterize the clustering properties of galaxies on the sky plane, in terms of the joint probability $\delta P=N^2[1+\omega(\theta)] \delta \Omega_1 \delta \Omega_2$ of finding two galaxies in the infinitesimal solid angles $\delta \Omega_1$ and $\delta \Omega_2$ separated by an angular distance $\theta$ given a surface density of objects $N$, with respect to what is expected for a Poissonian random distribution \citep{peebles80}.

  Usually, $\omega(\theta)$  is estimated by comparing the count of galaxy pairs with angular separation $\theta$ found in the observed field to the number of similar pairs obtained from a catalog of random and independently distributed objects. Several estimators have been designed to quantify this statistic, and in this work we use the estimator proposed by \cite{landy93}, which has the advantage of being unbiased and showing nearly Poisson variance:

\begin{equation}
\hat{\omega}(\theta)=\frac{\frac{DD(\theta)}{n_{dd}}-\frac{2DR(\theta)}{n_{dr}}+\frac{RR(\theta)}{n_{rr}}}{\frac{RR(\theta)}{n_{rr}}}
\end{equation}

Here, $DD(\theta)$ is the observed number of galaxy pairs with separations in the interval $[\theta,\theta\!+\!\delta \theta]$, $RR(\theta)$ is the analogous number obtained from a random catalog that completely imitates the field angular geometry, and $DR(\theta)$ is the amount of observed-random cross pairs with separations in the cited range. Each of these quantities is normalized by the total number of pairs in the sample, which correspond respectively to $n_{dd}=N_d(N_d-1)/2$, $n_{rr}=N_r(N_r-1)/2$ and $n_{dr}=N_rN_d$, with $N_d$ and $N_r$ the total number of objects in the observed and random catalogs.

We calculated $\hat{\omega}(\theta)$ in angular distance bins of constant logarithmic width $\Delta log(\theta)=0.15$. In order to produce random samples that replicate the survey geometry, we constructed angular masks over the K-band images used for galaxy detection, tracing the field borders and leaving out regions occupied by bright foreground stars or any image artifacts that forbid source detection. We then generated catalogs of 1000 random points located inside the angular mask, and calculated $\hat{\omega}(\theta)$ repeatedly over 25 different catalogs in order to amount to a total random sample $\sim100$ times larger than the observed one. We counted random, observed and crossed pairs in each field separately, and then combined the resulting $DD(\theta)$,$RR(\theta)$ and $DR(\theta)$ for each bin to produce the angular correlation function for all sBzK and pBzK galaxies found in the three fields of the MUSYC Wide NIR Survey. Since our three fields have practically identical areas and depths, the number of galaxies in each field is dominated by large scale structure rather than observational limits. Because of this we have taken the approach of calculating a single correlation function for the whole survey by counting all pairs in the three fields instead of computing a single $\omega(\theta)$ for each field and averaging the results. We consider our approach to be more statistically robust than the latter since low number of objects in an angular bin can translate into very noisy estimations. A full derivation of this approach of handling multiple fields to calculate $\omega(\theta)$ will be given by \cite{francke08}. The uncertainty in $\hat{\omega}(\theta)$ was estimated using a jackknife method, in which we divided each field into 25 smaller regions, and repeated the estimations of $\hat{\omega}(\theta)$ eliminating one sub-area at a time. In this way the standard deviation for each angular distance bin was obtained. For a detailed study on the robustness of the jackknife method refer to the Appendix on \cite{zehavi02}.

The angular correlation function is typically modeled by a power law of the form $\omega(\theta)=A_{\omega}\theta^{1-\gamma}$. The limited size of our sample does not allow an independent and significant measure of both the amplitude and the slope of $\omega(\theta)$, so in the rest of the analysis we assume a value of $\gamma=1.8$, consistent with slopes measured in faint and bright galaxy surveys \citep{zehavi02}. This also allows a direct comparison with previous work \citep{kong06, hayashi07}. Since we are estimating $\omega(\theta)$ in a finite region of the sky, we are affected by an uncertainty in the estimation of the background galaxy surface density \citep{peebles80, infante94} that is corrected by introducing a negative offset so that a more accurate parametrization of $\omega (\theta)$ is given by

\begin{equation}
\omega(\theta)=A_{\omega}(\theta^{1-\gamma}-C)
\end{equation}

This bias is know as the "integral constraint" and, as shown by \cite{roche99}, can be estimated numerically by

\begin{equation}
C=\frac{\sum{RR(\theta)\theta^{1-\gamma}}}{\sum{RR(\theta)}}
\end{equation}

which renders a value of $C=1.89$. Once we have applied this correction, we fit the amplitude through minimization of $\chi^2$. For fitting purposes we only consider separations in the $15''<\theta<0.1^{\circ}$ range. The lower limit corresponds to half a virial radius ($r_{200}$) of a halo of $\sim10^{13}M_{\odot}$ at z=2, in order to avoid the contribution of the ``one-halo'' term (i.e. clustering signal coming from galaxies that share DM halos). We experimented by setting larger lower limits for the bins to be considered for the fit (up to $\theta>40''$), and observed variations in the final values of the correlation lengths $r_0$ of the order of $0.1\sigma$, so we are confident of not being affected by effects introduced by galaxies inhabiting the same halos. The upper limit is taken in order to avoid spurious signal coming from border effects.

\begin{figure}[ht]
\begin{center}
\plotone{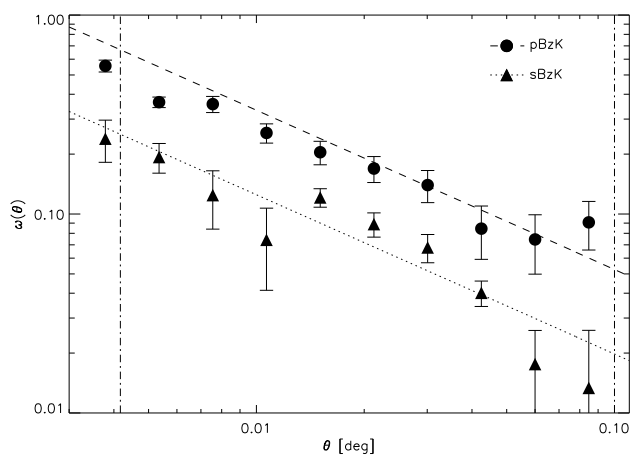}
\caption{Measured angular correlation function of sBzK (triangles) and pBzK (circles) galaxies in the MUSYC Wide NIR Survey. Best power-law fits for sBzK (dotted) and pBzK (dashed) galaxies are also shown. The vertical dotted-dashed lines corresponds to the $15''<\theta<0.1^{\circ}$ range used for the fitting.}
\label{fig-12}
\end{center}
\end{figure}

The measured 2-point angular correlation function of passive and star-forming BzK galaxies in the MUSYC Wide NIR sample, together with the best fitting power-laws are shown in Figure \ref{fig-12}. It is clear that there is a positive correlation signal for both sBzK and pBzK galaxies, and that passive BzK galaxies are more strongly clustered than star-forming BzK galaxies at z$\sim$2. Most points in the correlation function of both sBzK and pBzK are consistent with a pure power-law fit within the 2$\sigma$ uncertainties and all of them are consistent to a 3$\sigma$ level. Therefore we have not found any significant evidence of structure in $\omega(\theta)$ for any of the two populations. 

Table \ref{tbl-6} reports the best fitting values for the amplitudes of $\omega(\theta)$. For sBzK galaxies we obtain, $A_{\omega}=3.14\pm1.12$ and for pBzK we get $A_{\omega}=8.35\pm1.55$ (both in units of $10^{-3}$). These values are systematically lower but still in statistical agreement with the only previous estimate of the correlation function of bright $K<20$ BzK galaxies by \cite{kong06}. They obtained values of $A_{\omega}=4.95\pm1.69$ and $A_{\omega}=10.40\pm2.83$ for sBzK and pBzK galaxies respectively in the Deep3a-F field. The systematically lower amplitudes of the correlation functions together with the lower surface densities (see section 9.1) measured for both sBzK and pBzK galaxies in this work compared to \cite{kong06} indicate that the Deep3a-F field, 8 times smaller than the area covered in this work, show a level of clustering above the cosmic average, demonstrating once again the importance of large area surveys in order to overcome cosmic variance.

\subsubsection{Correlation Lengths, Bias and Dark Matter Halo Masses}

Under the assumption of a power-law form for $\omega(\theta)$, the 2-point spatial correlation function $\xi(r)$ also corresponds to a power-law with index $-\gamma$. Usually, $\xi(r)$ is parametrized in terms of the correlation length $r_0$ and the slope $\gamma$ as

\begin{equation}
\xi(r)=\left(\frac{r}{r_0}\right)^{-\gamma}
\end{equation}

The parameter $r_0$  can be calculated from the angular correlation function using the inverse Limber transformation \citep{limber53, peebles80}, which leads to the relation expressed by \cite{kovac07}

\begin{equation}
A_{\omega}= r_0^{\gamma}\sqrt{\pi} \frac{\Gamma(\frac{\gamma-1}{2})}{\Gamma(\frac{\gamma}{2})}\frac{\int_{0}^{\infty} F(z) D_A^{1-\gamma}(z) N_{corr}(z)^2 g(z) dz}{\biggl[\int_{0}^{\infty} N_{corr}(z) dz \biggr]^2}
\end{equation}
 
\noindent where $A_{\omega}$ is the amplitude of $\omega(\theta)$, $D_A(z)$ is the angular diameter distance, $g(z)$ is a cosmology dependent expression given by

\begin{equation}
g(z)=\frac{H_o}{c} \biggl[ \left(1+z\right)^2 \sqrt{1+ \Omega_M z +\Omega_{\Lambda} \left[\left(1+z\right)^{-2}-1\right]} \biggr]
\end{equation}

and $F(z)$ accounts for the redshift evolution of $\xi(r)$ described by $F(z)=(1+z)^{-(3+\epsilon)}$, where $\epsilon=-1.2$ in the case of fixed clustering in comoving coordinates, $\epsilon=0$ if the clustering is fixed in proper coordinates or $\epsilon=0.8$ according to the prediction of linear theory \citep{brainerd95}. In this work we assume the first case of constant clustering in comoving coordinates. Finally, $N_{corr}(z)$ corresponds to the redshift distribution of the studied population, which we have modeled as Gaussian distributions with the parameters reported in Table \ref{tbl-6} (see section 9.2).

We considered two sources of error in our $r_0$ calculations: (1) the uncertainty in the measurement of $A_{\omega}$ ($\sigma_{r_0,A_{\omega}}$), and (2) the uncertainty in the photometric redshifts used to construct the redshift distribution of both sBzK and pBzK galaxies ($\sigma_{r_0,\sigma_z}$). We estimated these two errors separately and added them in quadrature to obtain the uncertainty in the reported $r_0$ values. In the first case we used a Monte-Carlo approach, varying the $A_{\omega}$ values within their measured uncertainties using a normal distribution of errors, and then computed the dispersion of the resultant $r_0$ values calculated using a fixed $N_{corr}(z)$ with the parameters reported in Table \ref{tbl-6}. In order to estimate the errors coming from the uncertainty in the redshift distribution, we followed the approach of \cite{hayashi07} who showed that the Limber transformation is insensitive to changes in $z_o$, with the width $\sigma_z$ of the redshift distribution the dominant source of uncertainty. As above, we have estimated the uncertainty coming from $\sigma_z$ by varying it within its error (assumed Gaussian), but now fixing $A_{\omega}$. Table \ref{tbl-6} independently reports both uncertainties for the two populations, together with the measured correlation lengths.

For sBzK galaxies we measured $r_0=8.89\pm2.03$ Mpc, value that is much lower, although bearly consistent to a 1$\sigma$ level, than the $r_0=12.14^{+2.9}_{-3.3}$ Mpc ($8.5^{+2.0}_{-2.3}$ $h_{100}^{-1}$Mpc) measured by \cite{hayashi07} for $K<20$ sBzK galaxies in the Deep3a-F sample by \cite{kong06}. This is in agreement with the above evidence showing that the Deep3a-F field is denser and more clustered than the median field. We also measured $r_0=10.82\pm1.72$Mpc for pBzK galaxies. This is the first measurement of the correlation length of passive BzK galaxies ever reported.

In the context of the $\Lambda$CDM scenario, galaxies form by the cooling and condensation of baryonic gas in the cores of dark matter halos. The galaxy spatial correlation function is then associated with the auto-correlation function of DM halos and the spatial distribution of galaxies of a certain type is biased in the same way as that of their hosting halos with respect to the underlying mass distribution \citep{mo96}. In the context of the ellipsoidal collapse model extension of the Press-Schechter formalism by \cite{sheth01} they are related to first approximation by the linear bias. Following the methods of \cite{quadri07a} and \cite{francke08}, we estimated the galaxy effective bias adopting the following definition:

\begin{equation}
b_{eff}^2=\frac{\sigma^2_{8,gal}}{\sigma^2_{8,DM}(z)}
\end{equation}

where $\sigma^2_8$ corresponds to the variance (of galaxies or dark matter) in $8h_{100}^{-1}$Mpc radius spheres. We then estimated the minimum mass $M_{DH}^{min}$ of DM halos hosting galaxies in our sample, given by

\begin{equation}
b_{eff}=\frac{\int_{M_{DH}^{min}}^{\infty}b(M_{DH})n(M_{DH})dM_{DH}}{\int_{M_{DH}^{min}}^{\infty}n(M_{DH})dM_{DH}}
\end{equation}

where $b(M_{DH})$ is the bias parameter for halos of mass $M_{DH}$ taken from \cite{sheth01}, and $n(M_{DH})dM_{DH}$ is the halo mass function derived by \cite{sheth99}. We assumed the simplest case of one galaxy per halo. All the above expressions were evaluated at the center of the redshift distribution for each of the two populations studied, that is $z_o=1.58$ for pBzK and $z_o=1.78$ for sBzK, and obtained values of $b_{eff}=2.93^{+0.59}_{-0.60}$ for star-forming BzK, which correspond to a minimum halo mass of $M_{DH}^{min}=4\times 10^{12}$M$_{\odot}$, and  $b_{eff}=3.27^{+0.46}_{-0.47}$ for passive BzK galaxies, corresponding to $M_{DH}^{min}=9\times 10^{12}$M$_{\odot}$.

\begin{figure}[ht]
\begin{center}
\plotone{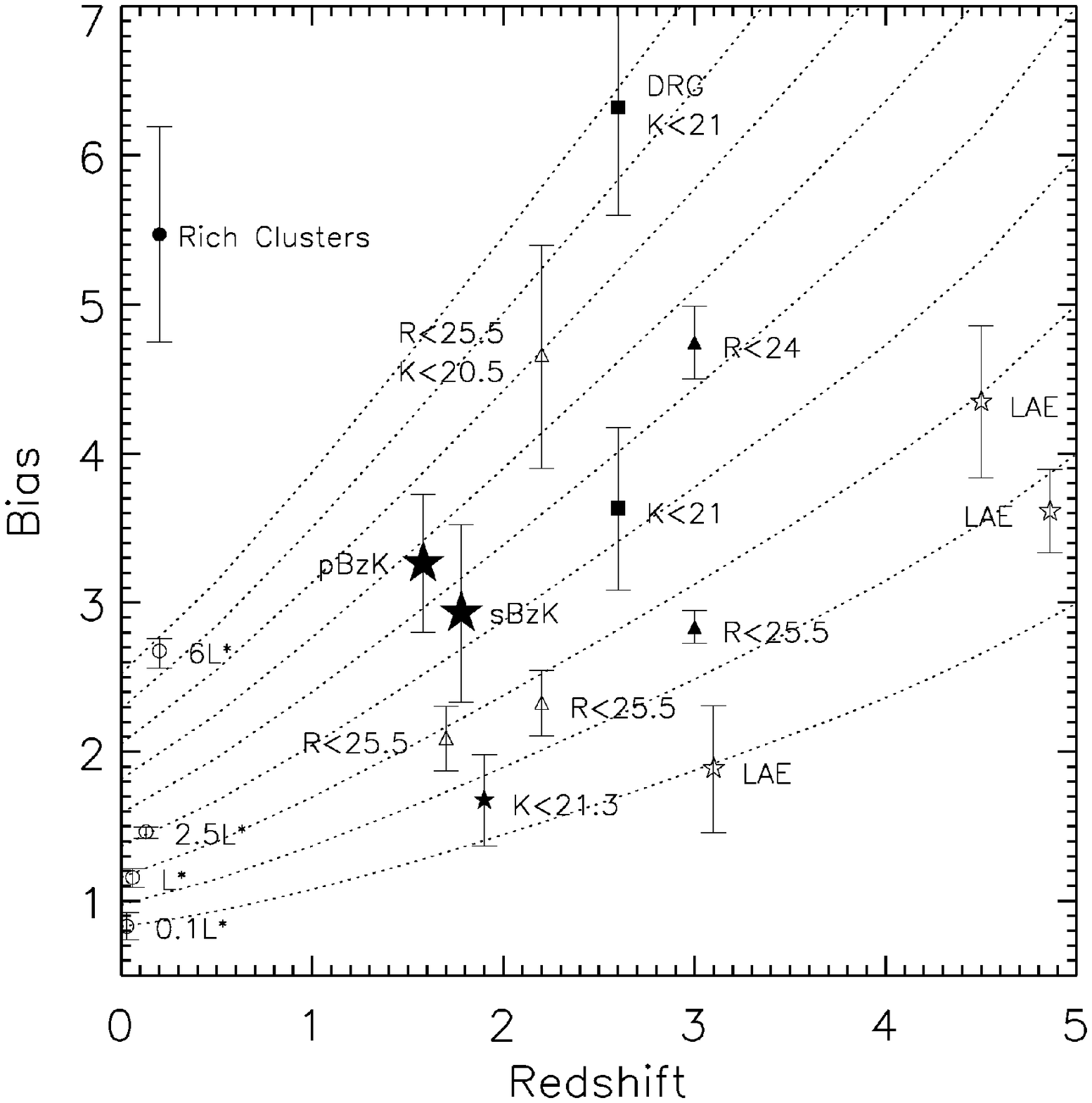}
\caption{Tracks of bias redshift evolution calculated using the ``halo merging'' model. Large filled stars show bias of $K<20$ sBzK and pBzK calculated in this work. Also presented are the bias of: faint sBzK galaxies from \cite{hayashi07} (small filled star); BM and BX galaxies with $\cal{R}<$$25.2$ from \cite{adelberger05a} and $K<20.5$ BX galaxies from \cite{adelberger05b} (open triangles); LBG at $z\sim3$ from \cite{lee06} (filled triangles), $K<21$ DRG and K-selected galaxies at $2.0<z<3.5$ from \cite{quadri07a} (squares), LAE at z=3.1, z=4.5 and z=4.86 (open stars, \cite{gawiser07, kovac07, ouchi03}, respectively), local SDSS galaxies from \cite{zehavi05} (open circles), and rich clusters from \cite{bahcall03} (filled circle).}
\label{fig-13}
\end{center}
\end{figure}

By comparing the bias of different populations of galaxies at different redshifts we can obtain insights on the evolution of these systems. Figure \ref{fig-13} shows tracks for the evolution of the bias with redshift together with measured bias factor values for different populations drawn from the literature. Previous work by \cite{quadri07a} and \cite{gawiser07} have presented similar plots using the ``no-merging'' model of \cite{fry96}. As dark matter becomes more clustered with time the bias factor of a biased population decreases. It is important to note that we expect mergers to have a more important role in higher density regions than in less dense regions where galaxies have a lower probablity of merging. Hence, including the effects of merging would translate into a steeper evolution of the bias with redshift. Therefore, the tracks calculated using the ``no-merging'' model only provide an upper limit for the bias factor of a given point at lower redshifts. In this work we present bias evolution tracks obtained from the ''halo merging model'' developed by \cite{bond91, bower91, lacey93} in which at any redshift halos have a given probability of merging into higher mass halos, so there is a defined mass distribution of descendant halos. For a given halo mass (bias) at a given redshift, the tracks follow the bias of the most likely descendant halo population, that is, the mode of the conditional mass distribution function evaluated at lower redshifts.

\begin{figure}[ht]
\begin{center}
\plotone{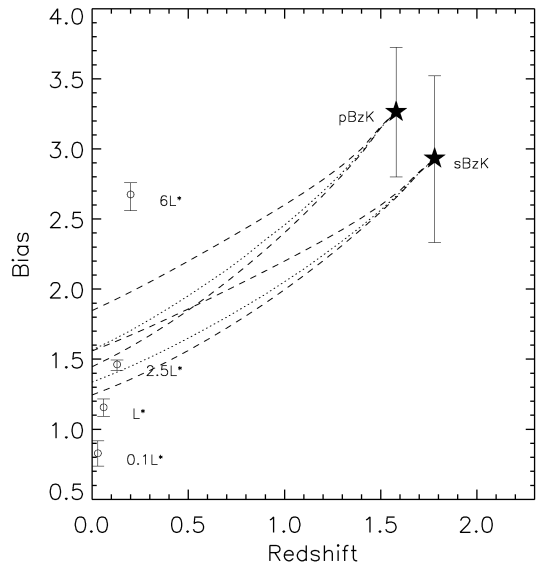}
\caption{Tracks of bias redshift evolution of $K<20$ pBzK and sBzK galaxies. Dotted lines show the bias of the most likely descendant halo population. Dashed lines mark the bias of halos whose likehood is reduced by a factor $exp(-1/2)$ with respect to the maximum likehood, in the case of a gaussian distribution of descendants this would correspond to $\pm 1\sigma$. The probability distribution of descendant halo masses is clearly non-symetric.}
\label{fig-14}
\end{center}
\end{figure}

It can be seen that although passive BzK galaxies in our sample show a higher bias than star-forming BzK galaxies, their clustering level is still consistent, and both populations inhabit halos in a similar mass range. Both sBzK and pBzK galaxies are consistent with being the descendant of the bright LBG population at $z\sim3$. Bright LBG have typical stellar masses of $\sim10^{10}-10^{11}$M$_{\odot}$ \citep{papovich01, iwata05} so, given the $\sim2$ Gy time elapsed between $z=3$ and $z=1.6$, they could easily be able to assemble enough mass to reach the typical masses of BzK galaxies given a reasonable SFR of $10-100$M$_{\odot}$yr$^{-1}$. High redshift LAE populations are not consistent with being the progenitors of BzK galaxies. It can also be seen that K bright UV selected galaxies at $z\sim2$ (the $K<20.5$ BX population) are more clustered than BzK galaxies, although the BX sample clustering measurement could suffer from cosmic variance \citep{adelberger05b}. In any case, the trend is consistent with what is observed at fainter magnitude limits, where faint sBzK galaxies show a lower clustering level than the BM/BX sample. On the other hand, as can be seen in Figure \ref{fig-14} the NIR bright BzK galaxies presented in this work are consistent with being the predecessors of massive early-type $>1$L$_{*}$ galaxies in the local universe usually present in groups and clusters, but are totally inconsistent with being associated with the progenitors of the central galaxies of rich clusters. The most likely descendants of $K$ bright sBzK and pBzK galaxies have bias factors of $1.3^{+0.2}_{-0.1}$ and $1.6^{+0.3}_{-0.1}$ at $z=0$.  

\section{Summary and Conclusions}

In this paper we have presented K-band imaging of two of the three $30'\times30'$ fields that conform the MUSYC Wide NIR Survey. We have given details of our data reduction procedures, as well as our source detection algorithms. By combining our data with the MUSYC UBVRIz optical imaging, we have constructed K-band selected source catalogs that reach a 5$\sigma$ point-source limiting depth of $K=20$. Catalogs are publicly available as part of the MUSYC Public Data Release, and present spatial and photometric information for the 16130 galaxies and 2891 stars brighter than $K=20$ detected in the 1030 and 1255 fields. We have also presented K-band differential number counts for galaxies in the two fields, which show an excellent agreement with previous K-selected samples.

By combining our data with the MUSYC K-selected catalog of the ECDF-S field, we have selected a sample of 2502 star-forming and 759 passively evolving BzK galaxies at $z\sim2$ over a very large area, allowing us to study the spatial distribution of these galaxies minimizing the effects of cosmic variance. We reported sky surface densities of $0.98\pm0.06$ arcmin$^{-2}$ for sBzK and of $0.30\pm0.10$ for pBzK galaxies. The large area surveyed over three non-contiguous fields allows the measurement of realistic uncertainties in the reported densities. We have found that the ECDF-S field is  underdense in passive galaxies at $z\sim2$ when compared to the mean for the complete survey, in agreement with previous studies on the density of DRG and K-selected massive galaxies. We have also confirmed the existence of the passively evolving early-type galaxy track at $z<1.4$ discovered by \cite{lane07}.

Using the methods calibrated by \cite{daddi04}, we estimated stellar masses of the BzK galaxies in our sample and showed that star-forming BzK galaxies have typical masses of $\sim1.0\times 10^{11}$M$_{\odot}$, while passive BzK objects have $\sim1.6\times 10^{11}$M$_{\odot}$. We also estimated the reddening and the SFR in our sample of star-forming BzK objects. sBzK galaxies show a median reddening of E(B-V)=0.47, implying that at least 55\% of them would be missed by UV continum based selection techniques. This is in excellent agreement with current studies on the overlap of galaxy populations at these redshifts \citep{reddy05, quadri07b, grazian07}. Star-forming BzK galaxies also showed large median SFR of $\simeq$230 M$_{\odot}$yr$^{-1}$, implying that bright $K<20$ sBzK can account for up to 30\% of the SFRD at $z\sim2$.

We also presented redshift distributions of both sBzK and pBzK, which show that both populations are selected over a narrow and well determined redshift range, with the pBzK population showing a narrower distribution than that of sBzK galaxies. We corrected these distributions to take into account the broadening introduced by errors in the photometric redshift estimation, and used them to deproject the angular correlation function of the two populations .

The 2-point angular correlation functions for pBzK and sBzK galaxies were presented, together with their best fits. We used the corrected redshift distributions to deproject the spatial correlation function from the angular one and estimated correlation lengths $r_0$ of $8.89\pm2.03$ Mpc and $10.82\pm1.72$ Mpc for sBzK and pBzK galaxies respectively, which translate into bias factor values of  $2.93^{+0.59}_{-0.60}$ and $3.27^{+0.46}_{-0.47}$ respectively. By comparing the effective bias of the spatial distribution of the two populations with that of DM halos, we estimated minimum halo masses of $4\times 10^{12}$M$_{\odot}$ and $9\times 10^{12}$M$_{\odot}$, for sBzK and pBzK respectively.

Finally, we compared the bias of different populations of galaxies at different redshifts, and concluded that NIR bright $K<20$ sBzK and pBzK galaxies are consistent with being the descendants of bright LBG at $z\sim3$, and the progenitors of local galaxies with $\la 1$L$_{*}$. This corresponds to the bright end of the luminosity function of present-day galaxies. The fact that K bright BzK galaxies evolve into very bright and massive systems is not surprising given their high stellar masses and large SFR at $z\sim2$. In the future, pushing the limits of K-band surveys to fainter magnitudes will allow us to detect large samples of the progenitors of current $<1$L$_{*}$ galaxies through BzK selection, overlapping these samples with the LAE population, opening a window to a more complete study of the ancestors of normal galaxies in the local universe.

\acknowledgements
We acknowledge the valuable support from FONDAP Centro de Astrof\'isica, FONDECYT projects \#1040423 and \#1040719, the Universidad de Chile and Yale University Astronomy Departments. We also thank the staff of Cerro Tololo Inter-American Observatory for their invaluable assistance with our observations. This research has made use of NASA's Astrophysics Data System. Finally we thank the referee for his helpful comments which helped to improve the quality of this work.

Facilities: CTIO(ISPI, MOSAIC II)

\clearpage

\begin{deluxetable}{llc}
\tabletypesize{\scriptsize}
\tablecaption{MUSYC Wide NIR Fields\label{tbl-1}}
\tablewidth{0pt}
\tablehead{
\colhead{Field} & \colhead{Equatorial Coord.} & 
\colhead{E(B-V)\tablenotemark{a}}
}
\startdata
 &   $\;\;\;\;\;\;\alpha\;\;\;\;\;\;\;\;\;\;
\;\;\;\;\;\delta$& mag\\
\tableline\\
SDSS1030+05 & 10:30:27.1 $\;$05:24:55 & 0.02\\ 
Cast1255 & 12:55:40.0 $\;$01:07:00 & 0.02\\  
E-CDFS & 03:32:29.0 -27:48:47 & 0.01\\  

\enddata
\tablenotetext{a}{\cite{schlegel98}}
\end{deluxetable}

\begin{deluxetable}{llccc}
\tabletypesize{\scriptsize}
\tablecaption{1030 and 1255 subfields K-band image properties\label{tbl-2}}
\tablewidth{0pt}
\tablehead{
\colhead{Field} & \colhead{Sub-field} & 
\colhead{Zero-point} & \colhead{FWHM\tablenotemark{a}} &
\colhead{5$\sigma$ depth\tablenotemark{b}} 
}
\startdata
 & & mag & arcsec & mag\\
\tableline
\\
SDSS1030+05 & C  & 21.48 $\pm$ 0.02 & 0.9 & 20.22 \\
            & NE & 22.01 $\pm$ 0.02 & 0.9 & 19.87 \\
            & E  & 21.99 $\pm$ 0.02 & 0.9 & 19.87 \\
            & SE & 22.14 $\pm$ 0.03 & 0.9 & 20.00 \\
            & S  & 21.94 $\pm$ 0.03 & 0.9 & 19.69 \\
            & SW & 22.02 $\pm$ 0.02 & 1.0 & 19.73 \\
            & W  & 22.05 $\pm$ 0.02 & 1.1 & 19.92 \\
            & NW & 21.98 $\pm$ 0.02 & 1.0 & 19.68 \\
            & N  & 22.03 $\pm$ 0.02 & 1.1 & 19.72 \\

Cast1255    & C  & 21.94 $\pm$ 0.02 & 0.9 & 19.86 \\ 
            & NE & 22.44 $\pm$ 0.03 & 0.9 & 19.84 \\ 
            & E  & 22.25 $\pm$ 0.03 & 1.0 & 19.72 \\ 
            & SE & 22.38 $\pm$ 0.01 & 0.9 & 20.00 \\ 
            & S  & 22.40 $\pm$ 0.03 & 0.9 & 19.77 \\ 
            & SW & 22.20 $\pm$ 0.03 & 1.0 & 19.77 \\ 
            & W  & 22.27 $\pm$ 0.03 & 0.9 & 19.89 \\  
            & NW & 22.40 $\pm$ 0.02 & 0.9 & 20.04 \\ 
            & N  & 22.28 $\pm$ 0.02 & 1.0 & 20.03 \\ 
\enddata
\tablenotetext{a}{Before PSF matching.}
\tablenotetext{b}{Vega magnitudes. After PSF matching}
\end{deluxetable}

\begin{deluxetable}{lcccc}
\tabletypesize{\scriptsize}
\tablecaption{MUSYC Optical Images of 1030 and 1255\label{tbl-3}}
\tablewidth{0pt}
\tablehead{
\colhead{Field} & \colhead{Filter} & 
\colhead{Zero-point} & \colhead{FWHM\tablenotemark{a}} &
\colhead{5$\sigma$ depth\tablenotemark{b}}
}
\startdata
 & & mag & arcsec & mag\\

\tableline
\\
SDSS1030+05 & U & 22.06 & 1.1 & 25.56\\ 
            & B & 24.97 & 1.0 & 25.91\\ 
            & V & 25.48 & 0.9 & 25.90\\ 
            & R & 25.75 & 0.9 & 25.81\\ 
            & I & 25.36 & 0.9 & 25.00\\ 
            & z & 24.36 & 1.0 & 23.75\\ 
Cast1255    & U & 23.32 & 1.2 & 25.62\\ 
            & B & 24.84 & 1.3 & 25.53\\ 
            & V & 24.90 & 1.1 & 25.42\\ 
            & R & 25.66 & 1.2 & 25.08\\ 
            & I & 24.94 & 1.1 & 24.00\\ 
            & z & 24.30 & 1.0 & 22.81\\ 
\enddata
\tablenotetext{a}{Before PSF matching.}
\tablenotetext{b}{AB magnitudes. After PSF matching}
\end{deluxetable}

\begin{deluxetable}{cccccc}
\tabletypesize{\scriptsize}
\tablecaption{Number of objects and sky densities per field\label{tbl-4}}
\tablewidth{0pt}
\tablehead{
\colhead{Field} & \colhead{Galaxies} & \colhead{sBzK} & \colhead{sBzK arcmin$^{-2}$}\tablenotemark{a} &\colhead{pBzK} & \colhead{pBzK arcmin$^{-2}$}\tablenotemark{a}
}
\startdata 
\\
1030 & 8015 & 758 & 0.92$\pm$0.03 & 441 & 0.50$\pm$0.02\\ 
1255 & 8115 & 911 & 1.10$\pm$0.04 & 165 & 0.20$\pm$0.02\\ 
ECDF-S & 8269 & 833 & 0.93$\pm$0.03 & 153 & 0.17$\pm$0.01\\ 
\\
Total & 24399 & 2502 & 0.98$\pm$0.06 & 759 & 0.30$\pm$0.10\\

\enddata
\tablenotetext{a}{Errors for individual fields correspond to Poisson errors. For the total survey we have included cosmic variance in the error estimation.}
\end{deluxetable}

\begin{deluxetable}{ccc}
\tabletypesize{\scriptsize}
\tablecaption{MUSYC BzK galaxies differential number counts\label{tbl-5}}
\tablewidth{0pt}
\tablehead{
\colhead{K bin center} & \colhead{sBzK} & 
\colhead{pBzK} 
}
\startdata
 mag & \multicolumn{2}{c}{log(N deg$^{-2}$ mag$^{-1}$)} \\

\tableline
\\
16.75 & 0.763 & - \\ 
17.25 & 0.939 & 1.064 \\ 
17.75 & 1.541 & 1.638 \\ 
18.25 & 2.275 & 2.313 \\ 
18.75 & 2.860 & 2.674 \\ 
19.25 & 3.300 & 2.864 \\ 
19.75 & 3.615 & 2.867 \\ 

\enddata
\end{deluxetable}

\begin{deluxetable}{lccccccccc}
\tabletypesize{\scriptsize}
\tablecaption{Clustering Properties of MUSYC BzK Galaxies\label{tbl-6}}
\tablehead{\colhead{Population} & 
\colhead{$N_{gal}$} & 
\colhead{$A\tablenotemark{a}_{\omega}$}  & 
\colhead{$z\tablenotemark{b}_o$} & 
\colhead{$\sigma\tablenotemark{b}_z$} & 
\colhead{$r_0$} & 
\colhead{$\sigma\tablenotemark{c}_{r_0,\sigma_z}$} & 
\colhead{$\sigma\tablenotemark{d}_{r_0,A_{\omega}}$} &
\colhead{Bias} &
\colhead{$M_{DH}^{min}$}
}
\startdata
  & & 10$^{-3}$ & & & $h_{70}^{-1}$Mpc & & & & M$_{\odot}$ \\

\tableline
\\

pBzK 	&  759	& $8.35\pm 1.55$ & 1.58 & $0.16\pm0.04$	& $10.82\pm1.72$ & 1.35 & 1.07 & $3.27^{+0.46}_{-0.47}$ & $1\times 10^{13}$\\
sBzK   	&  2502	& $3.14\pm 1.12$ & 1.78 & $0.31\pm0.06$ & $8.89\pm2.03$  & 1.00 & 1.77 & $2.93^{+0.59}_{-0.60}$ & $4\times 10^{12}$\\
\enddata
\tablenotetext{a}{Amplitude of the angular correlation function with slope parameter $\gamma=1.8$}
\tablenotetext{b}{Parameters of Gaussian fit to the redshift distribution $N_{corr}(z)$.}
\tablenotetext{c}{Dispersion in $r_0$ due to photometric redshift errors.}
\tablenotetext{d}{Dispersion in $r_0$ due to uncertainty in $A_{\omega}$.}

\end{deluxetable}

\clearpage

\end{document}